\documentclass[useAMS,usenatbib,fleqn]{mn2e}
\usepackage{amsmath,amssymb, aas_macros,times}% Include figure files
\usepackage{subcaption}
\usepackage{graphicx}
    \usepackage[flushleft]{threeparttable}
\usepackage[usenames,dvipsnames]{xcolor}
\usepackage{float}
\usepackage[utf8]{inputenc}
\usepackage{verbatim}
\usepackage{url}
\usepackage{caption}

\DeclareUnicodeCharacter{2212}{-}

\usepackage{mathptmx}
\newsavebox{\foobox}
\newcommand{\slantbox}[2][0]{\mbox{%
        \sbox{\foobox}{#2}%
        \hskip\wd\foobox
        \pdfsave
        \pdfsetmatrix{1 0 #1 1}%
        \llap{\usebox{\foobox}}%
        \pdfrestore
}}
\newcommand\unslant[2][-.18]{\slantbox[#1]{$#2$}}
\usepackage{soul}
\newcommand{\comm}[1]{}
\newcommand{\be}{\begin{equation}}
\newcommand{\ee}{\end{equation}}
\newcommand{\ba}{\begin{eqnarray}}
\newcommand{\ea}{\end{eqnarray}}

\newcommand{\bse}{\begin{subequations}}
\newcommand{\ese}{\end{subequations}}
\newcommand{\bwt}{\begin{widetext}}
\newcommand{\ewt}{\end{widetext}}
\usepackage{xcolor,cancel}
\begin{document}
\title[]{Study of maximum electron energy of sub-PeV pulsar wind nebulae by multiwavelength modelling}
\author[]{
Jagdish C.\ Joshi$^{1,2}$ \thanks{jagdish@aries.res.in},
Shuta J.\ Tanaka$^{3,4}$ \thanks{sjtanaka@phys.aoyama.ac.jp},
Luis Salvador Miranda$^{5}$ ,
Soebur Razzaque$^{2,6,7}$\\
$^1$ Aryabhatta Research Institute of Observational Sciences (ARIES), Manora Peak, Nainital 263001, India\\
$^{2}$ Centre for Astro-Particle Physics (CAPP) and Department of Physics, University of Johannesburg, PO Box 524, Auckland Park 2006, South Africa\\
$^{3}$ Department of Physical Sciences, Aoyama Gakuin University, 5-10-1 Fuchinobe, Sagamihara, Kanagawa 252-5258, Japan\\
$^{4}$ Graduate School of Engineering, Osaka University, 2-1 Yamadaoka, Suita, Osaka 565-0871, Japan \\
$^{5}$ Department of Physics, The Chinese University of Hong Kong, Shatin, Hong Kong China\\
$^{6}$ Department of Physics, The George Washington University, Washington, DC 20052, USA \\
$^{7}$ National Institute for Theoretical and Computational Sciences (NITheCS), South Africa \\ 
}

\bibliographystyle{mn2e}

%\begin{document}

\maketitle

\begin{abstract}
Recently, the Large High Altitude Air Shower Observatory (LHAASO) reported the discovery of 12 ultrahigh-energy (UHE; $\mathrm{\varepsilon} \ge 100$ TeV) gamma-ray sources located in the Galactic plane. A few of these UHE gamma-ray emitting regions are in spatial coincidence with pulsar wind nebulae (PWNe). We consider a sample of five sources; two of them are LHAASO sources (LHAASO J1908+0621 and LHAASO J2226+6057) and the remaining three are GeV-TeV gamma-ray emitters. In addition, X-rays, radio observations or upper limits are also available for these objects. We study multiwavelength radiation from these
sources by considering a PWN origin, where the emission is powered by spin-down luminosity of the associated pulsars. In this leptonic emission
model, the electron population is calculated at different times under the radiative (synchrotron and inverse-Compton) and adiabatic cooling. We also include the onset of the reverberation phase for the PWN, by assuming radially symmetric expansion. However, in this work, we find that multiwavelength emission can be interpreted before the onset of this phase. The maximum energy of the electrons based on the spectral fit is found to be above 0.1 PeV and close to 1 PeV. For LHAASO J2226+6057, using its observations in radio to UHE gamma-rays, we find that UHE gamma-rays can be interpreted using electrons with maximum energy of 1 PeV. We estimate the upper limits on the minimum
Lorentz factor of the electrons and it also infers the minimum value of the pair-multiplicity of charged pairs. 
\end{abstract}

%\date{\today}
\maketitle

\section{Introduction}
The PWN structure is energetically supported by the spin-down luminosity of the central pulsar and its composition is dominated by $e^{\pm}$ pair-plasma coupled with the magnetic field, as well as nuclei \citep{pacini1973ApJ249P, 1974MNRAS1671R,reynold1984Ap30R, Arons1994ApJS797A,bedna_prl2616B, 2003AnA827A, 2004AdSpR33456C, gaensler2006ARAG, volpi2008AnAV, kirk20091K,crabRPPh77f6901B,Kashiyama2017,Gelfand2017,Torres2017_ModNeb, amato2021Univ448A, halo22_NatAs199L}. A strong pulsar wind makes their nebula brighter in gamma-rays, and in general spin-down luminosity greater or equal to $4 \times 10^{36}$ erg/s is sufficient \citep{2004IAUS225G}.  In our Galaxy, PWNe is the  dominant class of very high energy (VHE; $100~ {\rm GeV}  \le \mathrm{\varepsilon} < 100$ TeV) gamma-ray sources, detected in the Galactic plane survey by the High Energy Stereoscopic System (H.E.S.S.) telescope \citep{abdalla2018AnA1H, HESS2018AnA612A2H}, also known as multiwavelength emitters \citep{reynolds2017SSRv175R}. \cite{Mattana2009ApJM} found that the production of TeV emission in PWNe is not correlated with the spin-down luminosity and the characteristic age of the pulsar. Further, for the VHE emission, the target photon field can be a combination of synchrotron photons, cosmic microwave background (CMB) photons, dust infrared (IR) and stellar photons \citep{tanaka2010ApJ1248T,torres2014JHEAp31T,2018zhu609A110Z,mares2021ApJ158M}. For older PWNe, the magnetic field is weaker, which makes them fainter in X-rays but due to the inverse-Compton (IC) scattering in the CMB, IR radiation they remain brighter in gamma-rays \citep{dejager2009arXiv4D}. In older PWNe, the gamma-ray emission is mostly due to the up-scattering of CMB photons or IR photons by relativistic electrons \citep{torres2014JHEAp31T}; and for younger ($t_{\rm age} < 300$~yr) PWNe, this emission is dominated by the upscattering of the synchrotron photons \citep{tanaka2011ApJT}. The detailed dynamical and radiative models of PWNe are useful to understand the physical parameters of the progenitor supernova (SN), energetics of the pulsar and its wind, properties of the surrounding environment, etc.\ \citep{Chevalier839C, gelfand2009ApJ2051G, torres2016MNRAS3868M, bandiera202051B}. The time-dependent leptonic spectral evolution model for PWNe can provide us with details about the electron population and magnetic field as functions of their age. The cooling of these electrons in the magnetic and radiation fields leads to a multiwavelength spectrum from radio to gamma-rays. For example, in the Crab Nebula, synchrotron self-Compton (SSC) mechanism and IC scattering off the CMB photons have been used to explain the  multiwavelength radiation \citep{tanaka2010ApJ1248T}. Their magnetic field evolution in time is also consistent with the rate of flux decrease in radio wavelengths.

%\noindent
In general, the multiwavelength emission from PWNe can be modelled using leptonic models \citep{esc_t_2008ApJ671210Z, tanaka2010ApJ1248T, tanaka2011ApJT, 2012MNRAS415M, 2013ApJ763L4T, torres2014JHEAp31T, 2018zhu609A110Z} or lepto-hadronic models \citep{atoyan1996MNRASA,2003AA0589B, LandH2009ApJ699L153Z,LiMNRAS408L80L}. The IceCube collaboration used the gamma-ray flux levels of 35 Galactic PWN sources and used stacking analysis to find the neutrino signal \citep{IC2020ApJ98117A}. They found neutrino flux in the TeV-PeV range is less than or equal to 4 $\%$ from these classes of objects. Recently, 12 UHE, Galactic gamma-ray sources have been discovered by the LHAASO and few of these gamma-ray sources are in spatial coincidence with some of the PWNe \citep{cao2021Natur33C, crpevlhaso}. In particular, the 1.1 PeV gamma-ray event was found to be associated with the Crab Nebula \citep{crpevlhaso} and the 1.4 PeV maximum energy photon was correlated with the Cygnus OB2 region \citep{cao2021Natur33C}. The radio to UHE gamma-ray emission from the Crab Nebula is consistent with the SSC+IC model and constrains the size of the electron pevatron in between 0.025 to 0.1 pc \citep{crpevlhaso}. They found that the luminosity in the PeV electrons is approximately 0.5 $\%$ of the pulsar's spin-down luminosity. This opens up a new domain of UHE gamma-ray Astronomy and powerful pulsars in our Galaxy play an important role in their origin \citep{albert2021ApJ27A}. The UHE gamma-ray detection is useful to test the theoretical models of electron acceleration in the PeV range \citep{giacinti2018ApJ18G,2021ApJ908L49B}. Further, PeV gamma-ray detection in PWNe can constrain accelerator size, the minimum acceleration rate, magnetic field, and the maximum Lorentz factor of electrons, etc.,\ \citep{crpevlhaso}.  Also, the pulsar wind can carry electrons of maximum energy and these are injected by the polar cap potential regions in the pulsars \citep{bucci_2011MNRAS381B}. The UHE gamma-ray sources and maximum photon energy detected from them provide some hint that particles are accelerating with maximum efficiency    \citep{deo_wilhelm_2022}. 

%\noindent
Here, we study multiwavelength emission from PWNe powered by their associated pulsars using the interactions between the non-thermal population of relativistic electrons with magnetic and radiation fields. The paper is organized as follows. In Section 2, we discuss the evolution of the PWN radius, cooling timescales that affects the electron distributions, magnetic field evolution, etc. In Section 3, we describe the PWN sources that have data or upper limits in radio, X-rays, and gamma-rays and perform modelling using a one-zone model. In Section 4, we discuss and conclude our results.

\section{The Model}

In this section, we have described our model, based on \cite{tanaka2010ApJ1248T} and further included the impact of SN reverse shock on the PWN radius inside a non-radiative supernova remnant (SNR) \citep{gelfand2009ApJ2051G}. The compression of the PWN radius can enhance the magnetic field and as a result, it affects the non-thermal radiation from relativistic electrons \citep{RC84, vander_rad_ev01,gelfand2009ApJ2051G,bandiera202051B}. These effects are most important if the PWN age is greater than 10 kyr \citep{halo22_NatAs199L}.

%\noindent
The spin-down luminosity of a pulsar at a given time can be estimated from the observed quantities, i.e., period $P$ of the pulsar, it's derivative $\dot{P}$ and moment of inertia $I$ of the neutron star (NS) \citep{gaensler2006ARAG}. Further, it enables a continuous supply of energy into particles and fields, that changes  according to the relation \citep{gaensler2006ARAG}
\begin{equation}
 L(t) = L_0 \left(1+ \frac{t}{\tau_0}\right)^{-(n+1)/(n-1)},
 \label{windL}
\end{equation}
\noindent
where $L_0$ is the initial spin-down luminosity and braking index $n$ is set equal to 3 for pulsars (PSRs) in our calculations. The total energy injected by a pulsar to its nebula in its lifetime approximately lies in the range of $\sim (1 - 5) \times 10^{-2}\,{E_{\rm SN}}$, where $E_{\rm SN} = 10^{51}$ erg is the SN kinetic energy \citep{gaensler2006ARAG,bucci_2011MNRAS381B}. The pulsar age $t_{\rm age}$, characteristic timescales, $\tau_c = P/2\dot{P}$ and $\tau_0$, are related by the relation \citep{tanaka2011ApJT}
\begin{equation}
\tau_c = \frac{n-1}{2} (\tau_0 + t_{\rm age}).
\label{eq_tage}
\end{equation}
As we have taken $n= 3$, in this case, the pulsar loses energy via its spin-down by the magnetic dipole radiation \citep{gaensler2006ARAG}. This value also implies that $\tau_c > t_{\rm age}$ to get a positive value of $\tau_0$. Note that the uncertainty in the age of the pulsar $t_{\rm age}$ affects the model parameters \citep{tanaka2013MNRAS5T}. Also, $\tau_0$ and $t_{\rm age}$ play important roles in shaping the spectral energy distribution (SED) of a PWN \citep{torres2014JHEAp31T}. As $\tau_c$ is known from pulsar observations and for a known or assumed value of  $t_{\rm age}$, we can estimate $\tau_0$. Further, these calculation depends on $n$, however, in our calculations we have taken $n =3$.

\subsection{Evolution of PWN Radius}
The radius of the PWN can be estimated analytically and is defined as \citep{Blondin2001ApJ06B}
\begin{equation}
\begin{split}
R_{\rm PWN}(t) \simeq 0.5~{\rm pc}~ \left(\frac{E_{\rm SN}}{10^{51}~\rm erg}\right)^{\frac{3}{10}} \left(\frac{M_{\rm ej}}{8 ~M_{\odot}}\right)^{-\frac{1}{2}} \\ \times \left(\frac{L_0}{10^{38}~{\rm erg/s}}\right)^{\frac{1}{5}} \left(\frac{t}{{\rm 500~yr}}\right)^{\frac{6}{5}}.
\end{split}
\end{equation}

\begin{figure}
\centering
\qquad
%\subfloat[label 2]{{\includegraphics[width=7cm]{lepton_pos.eps} }}%
\includegraphics[width=9 cm]{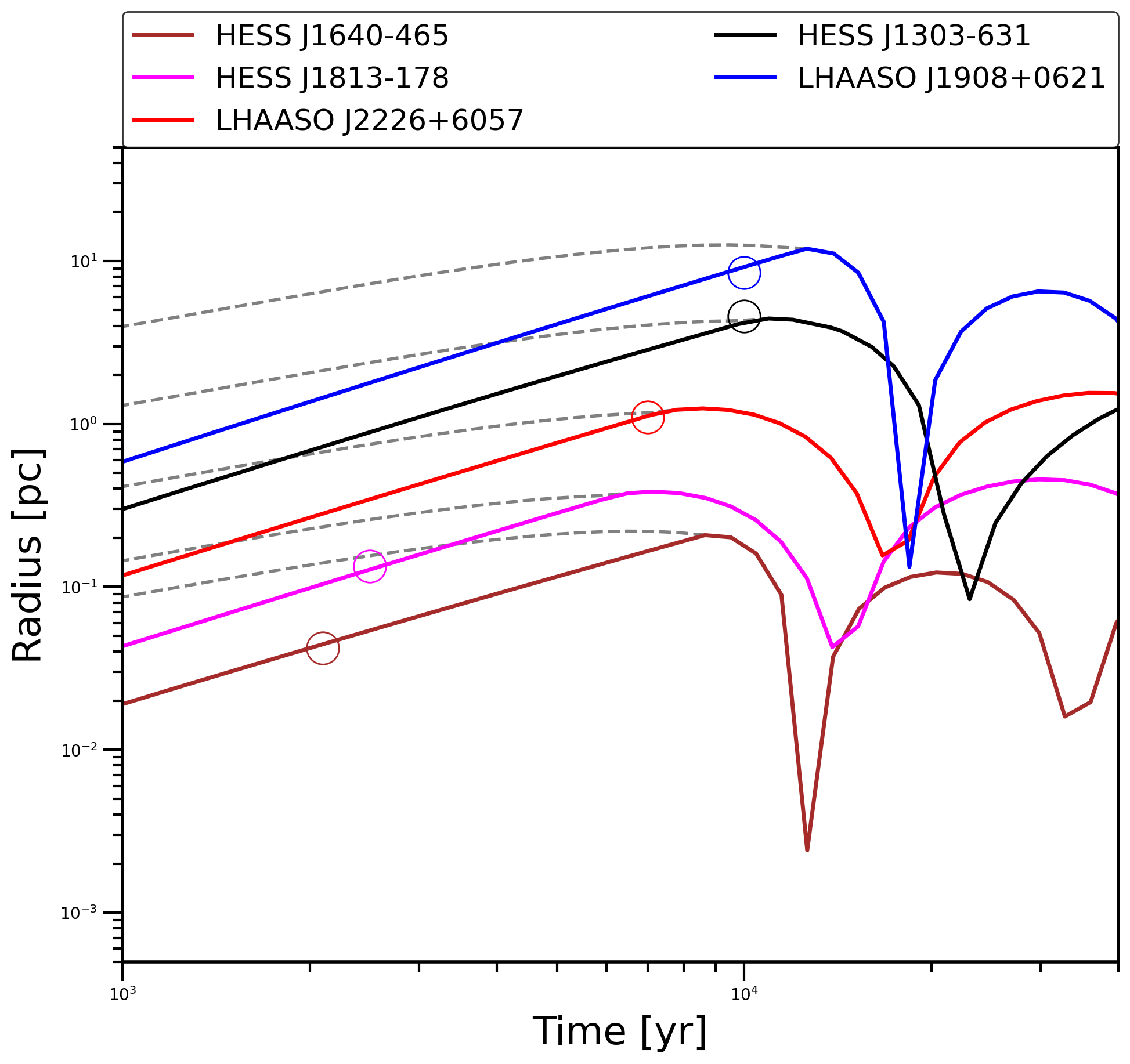}
\caption{The SN reverse shock radius (shown in grey color) and the evolution of the PWN radius. Once the SN reverse shock radius collides with the PWN radius the compression and later the re-expansion takes place. For better visibility of the curves, we have divided the radius values for HESS J1640-465, HESS J1813-178, LHAASO J2226+6057, and HESS J1303-631 by factors 50, 30,10, and 3 respectively. The circled point in each curve represents the radius at the current age of the PWN for which multiwavelength emission is calculated.}
\label{fig:radevl}
\end{figure}

\noindent
This analytical solution is only valid up to pulsar age $t_{\rm} \ll \tau_{0}$. $M_{\rm ej}$ is the amount of mass ejected  during the SN explosion. For the constant density $n_{\rm ISM}$ of the interstellar medium (ISM), the ejecta phase remains dominant up to a time $t_0 = 196 ~{\rm yr}~(M_{\rm ej}/M_{\odot})^{5/6} E_{\rm SN, 51}^{-1/2} n_{\rm ISM}^{-1/3}$ \citep{true1999ApJ299T}. Afterward, during the Sedov-Taylor (ST) phase, the SNR shock radius is given by $R_{\rm SNR} = 1.17 \times \left( E_{\rm SN} t^2/\rho \right)^{1/5}$, where the density $\rho = \mu_{\rm H} n_{\rm ISM}$ and the mean molecular weight of ISM is $\mu_H = 1.4 \times 1.67 \times 10^{-24}~{\rm g/cm^3}$ \citep{frank1992S}. Using these initial conditions we study the evolution of the PWN radius inside a non-radiative SNR \citep{gelfand2009ApJ2051G}. Based on this, the first compression starts when the PWN radius is equal to the SN reverse shock radius. After this stage, the pressure balance between the PWN volume and the ejecta material decides the PWN radius evolution. The details of the radius evolution are modelled based on the formalism discussed by  \cite{gelfand2009ApJ2051G} and the model is only valid up to the time when the ST phase ends. In Figure \ref{fig:radevl}, the dynamics of the PWN radius are shown w.r.t. the pulsar or PWN age. Initially, the PWN radius expands freely and after its collision with the SN reverse shock radius, the compression phase starts under the condition of sub-dominant pressure inside the PWN volume compared to the SN ejecta. In an opposite scenario, for later times, the PWN volume re-expands, and so on. The compression and re-expansion of the PWN volume are termed the reverberation phase. For our input parameters, in Table \ref{parameters}, we have listed the collision time $t_{\rm coll}$, after which compression starts. The onset of this phase dominates during the end of the ejecta phase or onward the ST phase. However, the details of the evolution also depend on the pulsar energetics and parameters. Based on Figure \ref{fig:radevl}, we also infer that for lower ejecta mass the compression starts early in time. We have kept the same value of SN energy and ISM density for all the objects, however, the onset of the compression phase is also sensitive to these parameters. The circled point in each curve represents the radius of the PWN at its current age $t_{\rm age}$, and in this work, we have investigated the multiwavelength radiation from all sources in the pre-contraction phase.

\begin{figure}
\centering
%\subfloat[label 1]{{\includegraphics[width=7cm]{leptons.eps} }}

%\includegraphics[width=7.cm]{elpop_1908.png}
\qquad
%\subfloat[label 2]{{\includegraphics[width=7cm]{lepton_pos.eps} }}%
\includegraphics[width=9 cm]{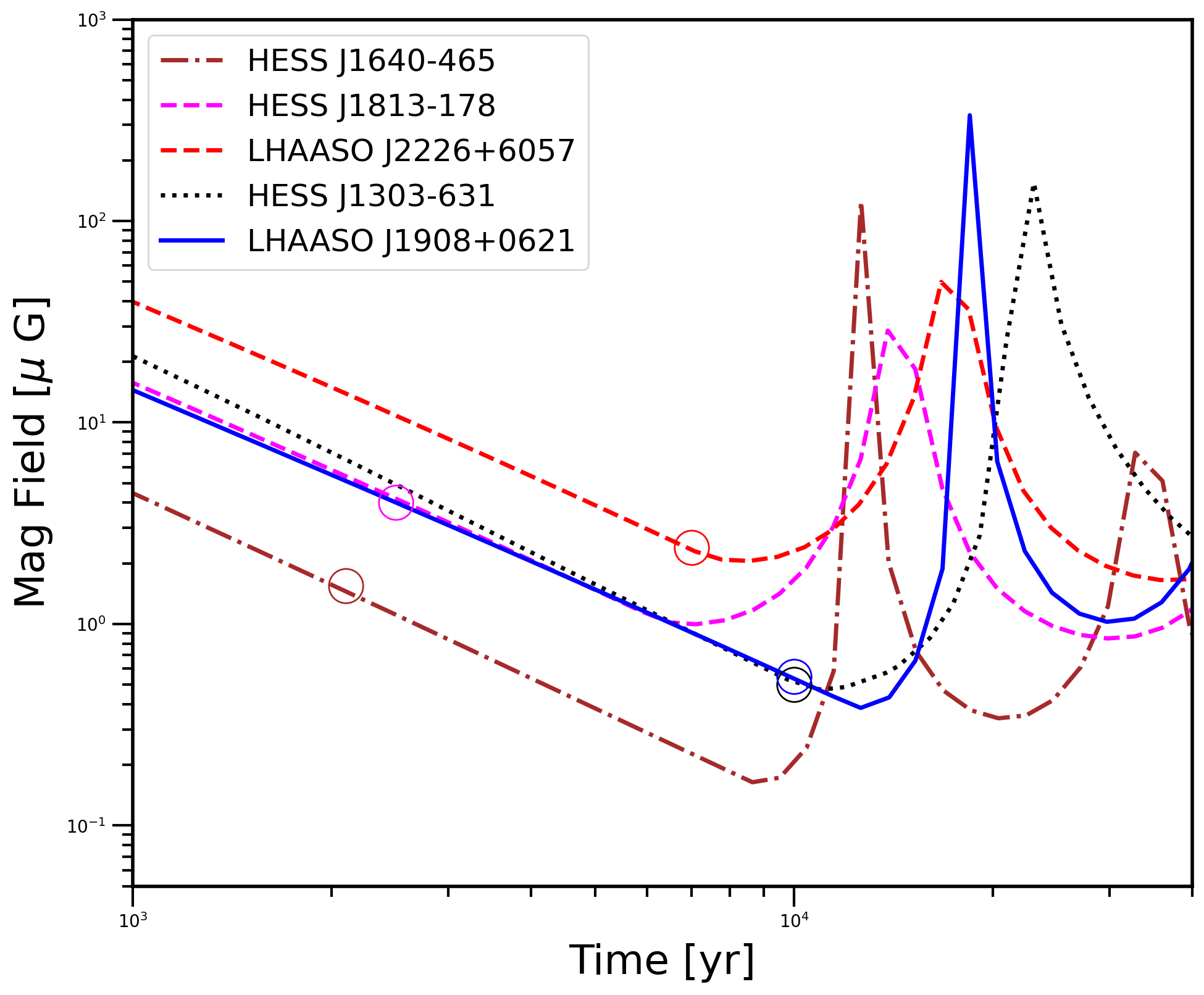}
\caption{The magnetic field inside the PWN and its evolution with time. Initially, the B-field decreases with the expansion of the pulsar wind and once the reverberation phase starts then the B-field undergoes expansion and re-compression phases, respectively. The circled point in each curve represents the magnetic field at the current age of the PWN for which multiwavelength emission is calculated. The magnetic field is estimated using equation 12. }
\label{fig:magevl}
\end{figure}

\subsection{Particle Distribution Under Radiative and Adiabatic Cooling}
\noindent
In our one-zone model, the time-dependent evolution of electron population $N(\gamma, t)$ at a time $t$, in the spherically expanding PWN  can be determined by their energy losses in the magnetic field, photon fields, and in addition via adiabatic losses. The particle distribution $N(\gamma,t)$ under energy losses $\dot{\gamma}(\gamma,t)$ and source term $Q(\gamma,t)$ can be found using the formalism discussed in  \cite{tanaka2010ApJ1248T}. The continuity equation for $N(\gamma,t)$ is defined as

\begin{equation}
 \frac{\partial N(\gamma, t)}{\partial t} + \frac{\partial }{\partial \gamma} [\dot{\gamma}(\gamma, t) N(\gamma,t)]   = Q(\gamma,t),
 \label{eq_cont}
\end{equation}
%+ \frac{N(\gamma, t)}{t_{\rm esc}}
\noindent
where $\gamma$ is the electron Lorentz factor. The second term on the left-hand side of equation \ref{eq_cont}, describes the cooling of relativistic electrons. We have considered the cooling rate due to adiabatic expansion $\dot{\gamma}_{\rm ad}(\gamma, t)$, synchrotron $ \dot{\gamma}_{\rm Sy}(\gamma, t )$ and IC $\dot{\gamma}_{\rm ic}(\gamma)$ scattering, respectively \citep{blu1970RvMP7B, tanaka2010ApJ1248T}. The adiabatic losses affect the low energy part of the electron distribution while the synchrotron and IC losses are important for larger values of the electron Lorentz factor \citep{tanaka2010ApJ1248T, vor2013ApJ7139V}.

We use the adiabatic cooling time as $t_{\rm ad} = |\gamma / \dot{\gamma}_{\rm ad}| = R_{\rm PWN}/V_{\rm PWN}$, and for our modelling this choice is okay as we are mostly in the expansion stage of the PWN radius \citep{2012MNRAS415M}. The values of PWN radius $R_{\rm PWN}$ and velocity $V_{\rm PWN}$ at the current age are listed in Table \ref{parameters} and same used for the estimation of $t_{\rm ad}$, as shown in Figures \ref{tA} to \ref{tC}. The electrons are cooled by synchrotron radiation in the magnetic field and the cooling rate for a single electron in the magnetic field is $\dot{\gamma}_{\rm Sy}(\gamma, t ) = -4 c \sigma_T  \gamma^2 U_{B}(t)/3m_ec^2 $, where $c$ is the speed of light, $\sigma_T$ is the Thomson cross section, $U_{B}$ is the magnetic energy density and $m_e$ is the electron mass. The synchrotron cooling time-scale is \citep{blu1970RvMP7B}

\begin{equation}
t_{\rm Sy} = \frac{6 \pi m_e c^2 }{c \sigma_T B^2 \gamma} = 2.5 \times 10^3 {\rm yr} \left(\frac{B(t)}{100~\mu{\rm G}}\right)^{-2} \left(\frac{\gamma}{10^6}\right)^{-1}.
\end{equation}

\noindent 
Further, the cooling rate of electrons in the target photon field including the Klein-Nishina (KN) scattering regime is $\dot{\gamma}_{\rm IC}(\gamma, t ) \approx -4 c \sigma_T  \gamma^2 U_{\rm ph}(t)/3m_ec^2(1 + 4 \gamma \epsilon_0/m_e c^2)^{3/2}$, and corresponding time-scale is  \citep{mode2005MNRAS.363..954M}

\begin{equation}
\begin{split}
 t_{\rm IC} = \frac{3}{4}\frac{m_e c^2 }{c \sigma_T U_{\rm ph}(t) \gamma}\left[1+\frac{4 \gamma \epsilon_0}{m_e c^2}\right]^{3/2} \\= 6.1 \times 10^5 {\rm yr} \left(\frac{U_{\rm ph}}{1~{\rm eV\,cm^{-3}}}\right)^{-1} \left(\frac{\gamma}{10^6}\right)^{-1} \left[1+\frac{4 \gamma \epsilon_0}{m_e c^2}\right]^{3/2}.
\end{split}
\end{equation}

\noindent
The target photons for the IC mechanism are synchrotron photons, CMB, IR photons
from dust grains and stellar optical photons. In the IC scattering, the Thomson regime is valid if the target photon of energy $\epsilon_0$ interacts with an electron with Lorentz factor $\gamma$ satisfying the condition $4 \gamma \epsilon_0/m_ec^2 \ll 1$ and later the KN effects are important when $4 \gamma \epsilon_0/m_ec^2 \gg 1$ \citep{jones1968PhR9J,blu1970RvMP7B,2021ChPhL38c9801F}. While interacting with the graybody photons with temperature $T$, KN effects become important when the energy of the relativistic electron is larger than $0.27 m_c^2 c^4/k_B T$ \citep{sch2010NJPhS}. Thus KN process reduces the energy loss for the electron and a harder spectrum for non-thermal electrons is expected.
\noindent
These electrons interact with the CMB radiation field with energy density $U_{\rm CMB} = 0.26 ~{\rm eV/cm^3}$ but the IR and stellar photon density depend on the source location in the Galaxy \citep{porter229P}. This should create a harder electron spectrum above the electron Lorentz factor $\gamma_{\rm KN} = 0.27 m_e c^2/k_B T$ \citep{sch2010NJPhS}, i.e., $6 \times 10^8, 8 \times 10^7$ and $3 \times 10^5$, for the pair-plasma interactions with the CMB, IR and stellar photons, respectively.

\noindent
The source term $Q(\gamma,t)$ is due to the pulsar wind that depends on time. In a phenomenological model, the broadband emission from PWN is generally explained using a broken power-law distribution of electrons. The broken power law  has a harder index $p_1$ in the range from 1 to 1.8, below the break Lorentz factor $\gamma_b$ and above it a softer index $p_2$ can take values in the range from 2 to 3.1 \citep{2013ApJ763L4T, tanaka2017ApJ8T, 2018zhu609A110Z}. 
\noindent
The time-dependent injected spectrum of electrons from the pulsar into the PWN can be either a power-law or a broken power-law type. 

\begin{align}
 Q(\gamma,t) & = Q_{0,e}
\begin{cases}
  \left(\frac{\gamma}{\gamma_b}\right)^{-p1},~~~~~~~  \gamma_{\rm min}< \gamma \le \gamma_b\\
  \left(\frac{\gamma}{\gamma_b}\right)^{-p2}, ~~~~~~~   \gamma_b < \gamma < \gamma_{\rm max}.
  \end{cases}
\end{align}

\noindent
The normalization $Q_{0, e}$ can be estimated using the relation,
\begin{equation}
\eta_e L(t) = \int_{\gamma_{\rm min}}^{\gamma_{\rm max}} Q(\gamma, t) \gamma m_e c^2 d\gamma,
\label{ener_ine}
\end{equation}

\noindent
where $\eta_e= L_e(t)/L(t)$ is the fraction of total energy into electrons. Further, we consider that $\eta_B = L_B(t)/L(t)$ is the fraction of total pulsar energy, available for the amplification of the magnetic energy. Approximately $10 \%$ of the pulsar spin-down energy radiates via pulsed emission \citep{vor2013ApJ7139V}.  These fractional parameters are used to estimate the value of the magnetization parameter $\sigma $, which is the ratio between magnetic energy flux and the particle energy flux, $\sigma = L_B(t)/(L_e(t) + L_A(t)) \approx \eta_B$ \citep{KCApJ283694K}. Here, $L_e(t) + L_A(t)$ represents the total pulsar energy distribution in $e^{\pm}$ pairs and nuclei of mass number $A$. In this work, we assume $L_A(t) = 0$ and hence all the injected energy is distributed in between the magnetic field and the $e^{\pm}$ pairs.  The value of $Q_{0,e}$ can be estimated by using the following expression

\begin{equation}
Q_{0,e} = \frac{\eta_e L_0}{m_e c^2} \left( 1 + \frac{t}{\tau_0} \right)^{-\frac{n+1}{n-1}} \left[\frac{\gamma_b^p \gamma_{\rm max}^{2-p}}{2-p} - \frac{\gamma_b^p \gamma_{\rm min}^{2-p}}{2-p}   \right]^{-1} .
\label{eqq0e}
\end{equation}

\noindent
While most of the particles inside the PWN are accelerated at the termination shock region but the polar cap potential is useful to scale the maximum energy of particles \citep{bucci_2011MNRAS381B}. The polar cap (PC) potential injected maximum energy of the particles is \citep{gold1969ApJ869G} 
\begin{equation}
 E_{\rm max, PC} \approx 6 \times 10^{12}~{\rm eV}~ \left(\frac{B_p}{10^{12}~ {\rm G}} \right) \left( \frac{R_{\rm NS}}{10~ {\rm km}} \right)^3 \left(\frac{P}{1~{\rm s}}\right)^{-2},
\end{equation}

\noindent
where $B_p \approx 3.2 \times 10^{19} (P \dot{P})^{1/2}~$G is the magnetic field on the surface of the NS, the radius of the NS is $R_{\rm NS} \approx 10$ km.  The second limit can be derived by comparing the synchrotron cooling time-scale of $e^{\pm}$ pair-plasma in the magnetic field $B$ of the termination shock with the acceleration time-scale $t_{\rm acc} =\pi \gamma m_e c^2 / e B c \eta_{\rm acc}$, i.e., time taken for completing half of a full gyration \citep{giacinti2018ApJ18G}. This is given by the expression

\begin{equation}
E_{\rm max, cool} = \left( \frac{6 m_e^2 c^4 e \eta_{\rm acc}}{\sigma_T B}\right)^{1/2} \approx 3.4 ~{\rm PeV}  \left(\frac{B(t)}{100~ \mu G}\right)^{-1/2} \left(\frac{\eta_{\rm acc} }{1}\right)^{1/2},
\end{equation}

\noindent
where $\eta_{\rm acc} \le 1$ is the acceleration efficiency parameter \citep{dejager1996ApJ3D}. We note that this is an upper limit on the maximum energy due to the cooling of particles, as effects due to Compton cooling are not considered. However, this is sufficient for our purpose of finding the possible reservoir for the injection of maximum energy of particles inside the PWN. The value of $\eta_{\rm acc} \le 1$ can be found by the maximum value of the synchrotron photon energy and $\eta_{\rm acc} = (\epsilon_{\rm max, Sy}/ 230~ {\rm MeV}$) \citep{amato2021Univ448A}, where $\epsilon_{\rm max, Sy} = h \nu_{\rm max}$, is listed in Table \ref{parameters}.  The break energy for a given $\gamma$ is calculated by using the expression $h\nu = 17.4 ~{\rm keV}~ (\gamma/10^8)^2 (B/100 ~{\mu \rm G})$ \citep{blu1970RvMP7B}.

\noindent
The magnetic energy of the PWN is supported by the spin-down luminosity $L(t)$ by an amount $\eta_B$. Also, adiabatic losses affect the total amount of magnetic energy available inside the PWN. We use the following equation to estimate the magnetic field, as also discussed earlier by \citep{torres2016MNRAS3868M},
\begin{equation}
\frac{dW_B}{dt} = \eta_B L(t) - \frac{W_B(t)}{R_{\rm PWN}(t)}\frac{dR_{\rm PWN}(t)}{dt},
\label{eq:bfield}
\end{equation}

\noindent
where $W_B = B^2R_{\rm PWN}^3/6$. The evolution of the magnetic field is shown in Figure~\ref{fig:magevl} and during the compression phase magnetic field gets amplified.

The average energy of the particles $E_w \approx \Gamma_w m_e c^2$ inside the PWN is approximately $\gamma_b m_e c^2 (\gamma_b/\gamma_{\rm min})^{-p_1 +1}$. We have also estimated the pair multiplicity $\kappa$ that is defined as $\kappa = \int_{\gamma_{\rm min}}^{\gamma_{\rm max}} Q(\gamma,t) d \gamma /\dot{n}_{\rm GJ}$, where $\dot{n}_{\rm GJ} = (cI \Omega \dot{\Omega}/e^2)^{1/2}$ is the Goldreich–Julian number flux \citep{gold1969ApJ869G}.  The estimated values are listed in Table \ref{parameters} and provides insights on the pulsar environment.

\begin{table*}
\centering
\caption{Model parameters for the PWN based on the observations and spectral fitting. The symbols that are based on the observations are pulsar period $P$, period derivative $\dot{P}$, distance to the pulsar $d$, characteristic age $\tau_c$, surface magnetic field of the NS $B_p$, current spin-down luminosity $\dot{E}$ respectively and wavelength dependent size of the PWN, if available from observations. We have assumed a value for PSR age $t_{\rm age}$, a fixed value for the input parameters; braking index $n$, interstellar medium gas density $n_{\rm ISM}$, SN kinetic energy $E_{\rm SN}$, IR energy density $U_{\rm IR}$ and optical energy density $U_{\rm Opt}$. However, for HESS J1640-465, a large value of energy density in optical photons is motivated from recent observations.
We fit the pulsar or PWN age $t_{\rm age}$,  ejecta mass $M_{\rm ej}$ of the progenitor star,  spectral index $p_1$, $p_2$, electron distribution with Lorentz factors $\gamma_{\rm min}, \gamma_{\rm b}, \gamma_{\rm max}$ and fractions of the total energy in electrons and magnetic field $\eta_e$ and $\eta_B$ based on the multiwavelength observations. The derived parameters from the above information are: pulsar spin-down timescale $\tau_0$, initial spin-down luminosity of the pulsar $L_0$, PWN magnetic field $B$,  radius $R_{\rm PWN}$ and velocity $V_{\rm PWN}$, etc., at $t_{\rm age}$. $t_{\rm coll}$ is the collision time of the PWN radius with the SN reverse shock. Further, the value of average Lorentz factor $\Gamma_w$, pair multiplicity $\kappa$ is estimated. The last four parameters are electron acceleration efficiency $\eta_{\rm acc}$ based on maximum synchrotron energy and corresponding maximum electron Lorentz factor allowed by the synchrotron cooling $\gamma_{\rm max, cool}$ and $\gamma_{\rm max, PC}$ is the allowed electron Lorentz factor due to the pulsar injection.}
  \begin{tabular}{llllllllll}
  \hline
  Model parameters       &     LHAASO J1908+0621  & LHAASO J2226+6057 & HESS J1640-465 &HESS J1813-178 & HESS J1303-631\\
         &   (PSR J1907+0602)  & (PSR J2229+6114) & (PSR J1640-4631) & (PSR J1813-1749 ) & (PSR J1301-6305) \\
 \hline
 &   From Past Observations &   \\
 \hline
$P$ (ms)   &   106.6[1]    & 51.6[3] &  206 [5]    & 44.7[8]  &  184 [10]                            \\
$\dot{P}$ (s/s) &  $87.3 \times 10^{-15}$[1] & $78.3 \times 10^{-15}$[3] &$9.758 \times 10^{-13} $[5] & $1.265 \times 10^{-13} $[8] & $ 2.65 \times 10^{-13}[10] $  \\
$d$ [kpc] & $3.2 \pm 0.6$ [2] & 3 [4] & 10 [6] & 6.2  [9] & 6.6 [11]\\ 
$\tau_c$ (kyr)        &   19.4   & 10.5 & 3.1  & 5.6 & 11\\
$B_p$ ($10^{12}~$G)        &   3   & 2 & 14  & 2.4 & 7\\
$\dot{E}$[erg/s] &  $2.84 \times 10^{36}$[1] & $ 2.25 \times 10^{37}$[3]  & $4.4 \times 10^{36}$[5] & $ 6.8 \times 10^{37}$[8]  &  $1.7 \times 10^{36} $[10]  &              \\
Radio: Size   &   No counterpart [12]     &  $200^{''}$ (3 pc)[16] & $8^{'}$ (23 pc)[19]      &   $3^{'}$ (5.4 pc)[22]   & No counterpart [25] \\
Xray:Size        & $12^{''}$ 0.2 pc [13]     & $200^{''}$ (3 pc)[17] & $1.2^{'} $ (3.5 pc) [20]   & $80^{''}$ (2.4 pc) [23] & $2^{'}$ (3.8 pc) [26] \\
VHE(HESS):Size        &   $0.34^{\circ}$ (19 pc) [14]  &  & $12^{'}$ (35 pc) [21]   & $(2.2 \pm 0.4)^{'}$ (4 pc) [24] &  $(0.16 \pm 0.02)^{\circ} $ (18.4 pc) [27]\\
VHE(VERITAS):Size        &      & $0.27^{\circ}$ (14 pc)[18] &   & & \\
UHE (LHAASO):Size &  $0.45^{\circ}$ (25 pc)[15] & $0.49^{\circ}$ (25.6 pc) [15] &  &   &   &\\
\hline
 &   Assumed Parameters &   \\
\hline
$t_{\rm age}$ [kyr] & 10 & 7  & 2.1 & 2.5 & 10\\
$n$  & 3 & 3 & 3 & 3 & 3&\\
$n_{\rm ISM}$ [$\rm{cm^{-3}}$] & 0.1 & 0.1 & 0.1 & 0.1 & 0.1&\\
$E_{\rm SN}$ [$10^{51} \rm{erg}$] & 1 & 1 & 1 & 1 & 1&\\
$U_{\rm IR} [{\rm eV/cm^3}]$ & 0.2 & 0.2  & 0.2 &  0.2 & 0.2\\ 
$U_{\rm Opt} [{\rm eV/cm^3}]$ & 0.2& 0.2  & 1300 [28]  & 0.2  & 0.2\\ 
\hline
&   Fitted Parameters &   \\
\hline
 $M_{\rm ej}$ [$M_{\odot}$] & 12 & 10 & 8 & 8 & 12 &\\
$p_1, p_2$ & 1.3,2.32 &  1.9, 2.48 & 1.5, 2.1 & 1.5,2.15 & 1.3,2.1\\
$\gamma_{\rm min}, \gamma_b, \gamma_{\rm max} $ & ($5 \times 10^{2}, 8 \times 10^{5}, 1.4 \times 10^{9}$) &  ($2 \times 10^2, 10^{4}, 1.8 \times 10^{9}$) & ($10^2, 10^{5}, 1.2 \times 10^{9}$) & ($4 \times 10^{2}, 10^{4}, 1.2 \times 10^{9}$)&  ($10^{4}, 10^{5}, 6 \times 10^{8}$) \\
$\eta_e, \eta_B $ & ($\sim 1$, $6 \times 10^{-4}$) &  ($\sim 1$, $2.5 \times  10^{-3}$) & ($\sim 1$, $10^{-4}$) & ($\sim 1$, $5 \times 10^{-4}$) & ($\sim 1$, $ 5 \times  10^{-4}$)\\
\hline
&   Derived Parameters  & &    \\
\hline
$\tau_0$ [kyr] & 9.4 & 3.5 & 1 & 3.1 & 1& \\ 
$L_0$ [erg/s] & $1.2 \times 10^{37}$ & $2.0 \times 10^{38}$   &$ 4.2 \times 10^{37}$ &$ 2.2 \times 10^{38}$ &$ 2.1 \times 10^{38}$ &\\ 
$B(t_{\rm age}) [\mu$G]   & 0.55  & 2.4  & 1.6 & 4 &  0.5 &\\
$R_{\rm PWN}(t_{\rm age})$[pc]       & 8.5   &  11  &  2.1  & 4  &  13.8  &  \\
$V_{\rm PWN}(t_{\rm age})$[km/s]       & 1034  &  1750 &  1151 & 1733 &  1297 &  \\
$t_{\rm coll}$[kyr]       & 12.4  &  7.3 &  8.7 & 6.3 &  10.3 &  \\
$\Gamma_{w} $ & $8.7 \times 10^4$ & $1.6 \times 10^2$ & $3.2 \times 10^3$ & $2 \times 10^3$ & $5 \times 10^4$  &  \\
$\kappa $ & $6.5 \times 10^4$& $5.4 \times 10^7$ & $2.2 \times 10^6$ & $1.5 \times 10^7$ & $9 \times 10^4$  &  \\
$\eta_{\rm acc} $ & $7.4 \times 10^{-5}$ & $6 \times 10^{-4}$ & $1.7 \times 10^{-4}$ & $4.3 \times 10^{-4}$ & $1.4 \times 10^{-5}$ &  \\
$\varepsilon_{\rm max, Sy}$ [eV]       & $1.7 \times 10^4$ &    $1.3 \times 10^5$   & $4 \times 10^{4}$ &$10^5$ &  $3.1 \times 10^3$ &\\
$\gamma_{\rm max, PC} $ & $3.2 \times 10^{9}$ & $9 \times 10^{9}$  & $4 \times 10^{9}$   &$1.4 \times 10^{10}$  &$2.5 \times 10^9$  &  \\
$\gamma_{\rm max, cool} $ & $8.1 \times 10^{8}$ & $ 10^{9}$  & $7 \times 10^8$   & $7 \times 10^8$ &$3.5 \times 10^8$  &  \\
\hline
\end{tabular}
\label{parameters}
\begin{tablenotes}
      \small
      \item {\bf References:} [1]\cite{abdo2010ApJS460A}, [2]\cite{abdo2010ApJ64A}, [3]\cite{abdo2010ApJS460A}, [4]\cite{halpern2001ApJ125H}, [5]\cite{gott2014ApJ155G},  [6]\cite{Lemiere2009ApJ9L},  [7] \cite{2016819L..16A},  [8]\cite{halpern2012ApJ14H}, [9]\cite{yao2017ApJ29Y} [10] \cite{manches2005AJ129M}, [11]\cite{cordes2002astrC}. [12] \cite{duvi2020MNRAS5732D}, [13] \cite{abdo2010ApJ64A}, [14] \cite{hess1908src_AnA723A},  [15] \cite{cao2021Natur33C} [16] \cite{halp2001ApJ23H}, [17] \cite{halpern2001ApJ125H}, [18] \cite{acciar2009ApJ6A}, [19] \cite{Whit1996A9W},  [20] \cite{gott2014ApJ155G}, [21] \cite{abram2014ApJss}, [22] \cite{brogan2005ApJ5B}, [23] \cite{uber2005ApJ109U,funk2007AnAF,helfand1297H}, [24] \cite{disc_2006ApJ777A},  [25] \cite{sushch2017AnAS}, [26] \cite{hess2012AnA46H}, [27] \cite{ahar2005AnA3A}, [28] \cite{mares2021ApJ158M}. 
\end{tablenotes}
\end{table*}

\section{multiwavelength Emission Modelling under Adiabatic and Radiative Cooling}

\noindent
In this Section, we describe source properties, their multiwavelength radiation, and their known interpretation and then we discuss the source parameters based on our modelling. The criteria for the source selection is, the distances should be known, and most plausibly the X-ray and radio observations or upper limits are available. In the list of five sources we have studied, two of them are detected by the LHAASO detector in the TeV-PeV band. In the gamma-ray band, we have also included data from the Fermi-Large Area Telescope (LAT), H.E.S.S., Very Energetic Radiation Imaging Telescope Array System (VERITAS),  High Altitude Water Cherenkov Observatory (HAWC) and MILAGRO, if available.

\noindent
Based on our calculation, we have estimated the population of non-thermal electrons at different epochs. These electrons radiate from radio to UHE gamma-rays and these emissions are calculated using the python package NAIMA \citep{naima}. The synchrotron intensity is calculated using the formalism discussed in \cite{2010PhRvD82d3002A, baring1999ApJ311B}. Further, the IC of synchrotron and thermal photons is calculated using the formalism discussed in \cite{khangul2014ApJK}. We evolve the electron population for each object up to its current age and the electron population inside the PWN at four random epochs is shown in the top panel of Figures 3 $-$ 7. In the respective bottom panels, we have shown the spectral fit to the observational data at the current age of the PWN.  The breaks in the SED are either due to the injection spectrum of electrons $\gamma_{\rm min}, \gamma_b, \gamma_{\rm max}$ or the cooling break, defined as $\gamma_c(t) \sim 2.45 \times 10^6 (t/1{\rm kyr})^{-1} (B(t)/ 100 ~{\mu \rm G})^{-2}$ \citep{tanaka2011ApJT}.

\subsection{LHAASO J1908+0621}

LHAASO J1908+0621 is one of the Galactic UHE gamma-ray sources reported by the LHAASO collaboration with its possible connection to SNR G40.5-0.5, PSR J1907+0602 and PSR J1907+0631 \citep{cao2021Natur33C}. This source was also associated with the HAWC detected UHE gamma-ray source eHWC J1907+063 \citep{abey2020PhRvLA}. The spin-down luminosity of PSR 1907+0602  was sufficient to support the TeV emission of MGRO J1908+06 \citep{abdo2010ApJ64A} and associated HESS detection of J1908+063 \citep{hess1908src_AnA723A}. In the same region, the extended emission was reported by VERITAS; VER J1907+062 and its radio emission properties are investigated in detail and no radio counterpart was observed \citep{duvi2020MNRAS5732D}. Recently, \cite{crestan2021MNRAS09C} derived the X-ray upper limits for this source using the {\it XMM-Newton} observations. The emission from this spatial region has been studied in the case of MGRO J1908+06 as a Galactic pevatron for which the gamma-ray spectrum is harder above 100 TeV \citep{crestan2021MNRAS09C,Li2021ApJ33L}. Further, a PWN origin of the UHE gamma-rays has been investigated by \cite{bre5296B, crestan2021MNRAS09C,Li2021ApJ33L}.

We have also used our comprehensive model by considering the role of PSR 1907+0602 in explaining the multiwavelength emission. We have chosen this PSR compared to PSR 1907+0631 (another pulsar in the same region), due to its higher ($\sim 5$ times) current spin-down luminosity. The distance $d$ to this object, the current period $P$ and the period derivative $\dot{P}$ is $3.2 \pm 0.6$ kpc, 106 ms and $87.3 \times 10^{-15}$, respectively \citep{abdo2010ApJ64A}. Using the values of the period and its derivative, the corresponding value of $\tau_c$ is $\sim 19.4 $ kyr. The source is located at $(l, b) = [40.49^{\circ},-0.81^{\circ}]$ or at $(R,z) = [6.41, -0.05]$ kpc (by taking the distance to the Galactic centre 8.3 kpc) \citep{abdosecfc2013ApJ}. If we consider the age of the pulsar equal to the age of the SNR G40.5-0.5, that is found to be in between 25 to 40 kyr \citep{downes1980AnA47D}, then from equation \ref{eq_tage}, a negative value of $\tau_0$ is obtained. However, this issue can be resolved by taking a lower value of PWN age. To interpret the multiwavelength observations we have taken a fiducial value $t_{\rm age} = 10$ kyr. The emitting regions, constrained by the observations are $12^{''}$ in X-rays \citep{abdo2010ApJ64A}, the VHE size is $0.34^{\circ}$ \citep{hess1908src_AnA723A} and the UHE size of $0.45^{\circ}$ \citep{cao2021Natur33C}. As seen from the radius evolution curve for LHAASO 1908+0621 in Figure \ref{fig:radevl}, the radius is in the pre-compression phase and its value is approximately $R_{\rm PWN} = 8.5$ pc at the current age as shown by the blue circle in Figure \ref{fig:radevl}. That is greater than the X-ray size of the nebula but lower by a factor of 2 to 3 compared to VHE and UHE gamma-ray emitting regions. We will discuss in section 4, the possible reasons for the difference in the size between the model and observation.

\begin{figure}
\centering
%\subfloat[label 1]{a{\includegraphics[width=7cm]{leptons.eps} }}
\includegraphics[width=8 cm]{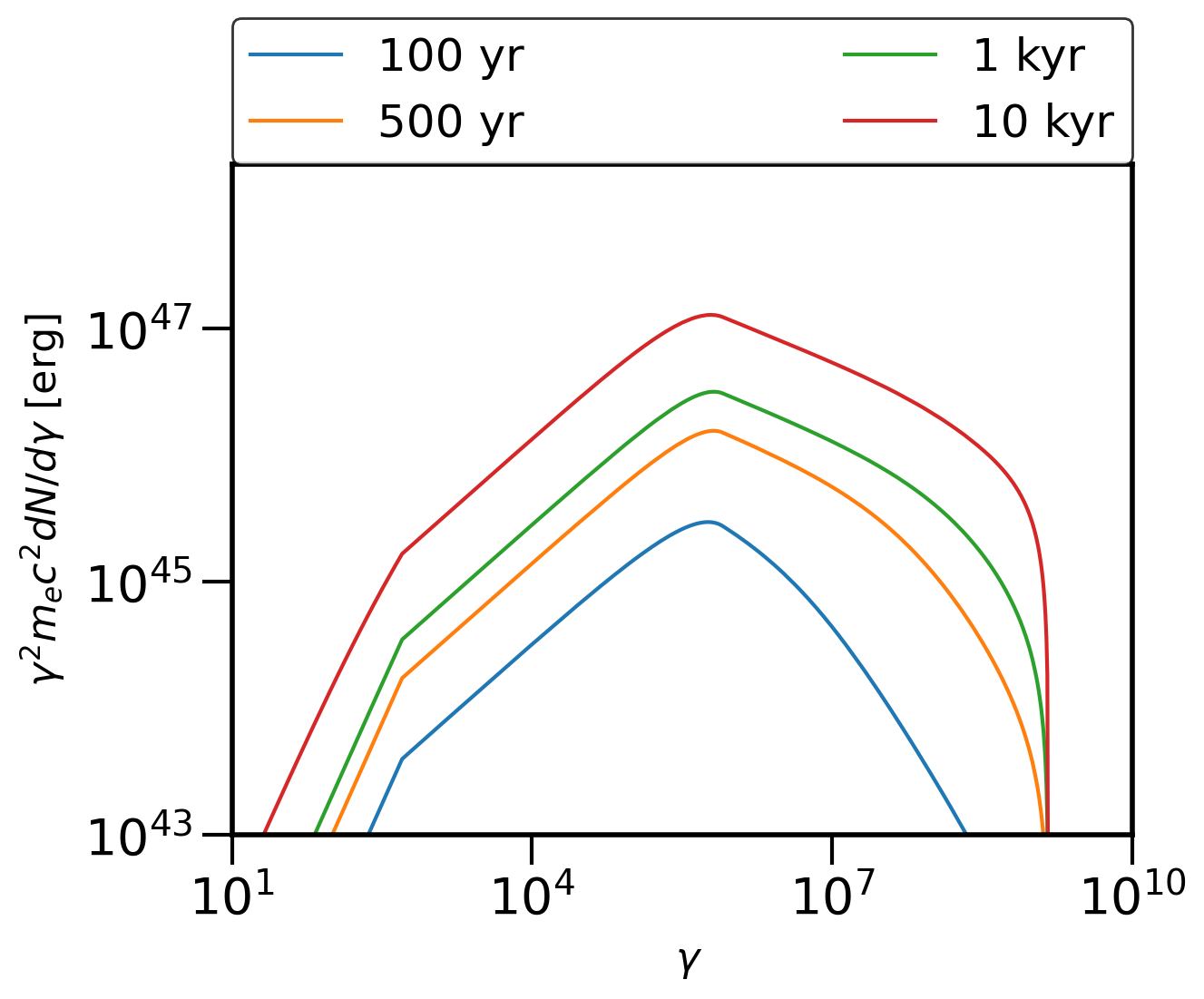}
\qquad
\includegraphics[width= 8 cm]{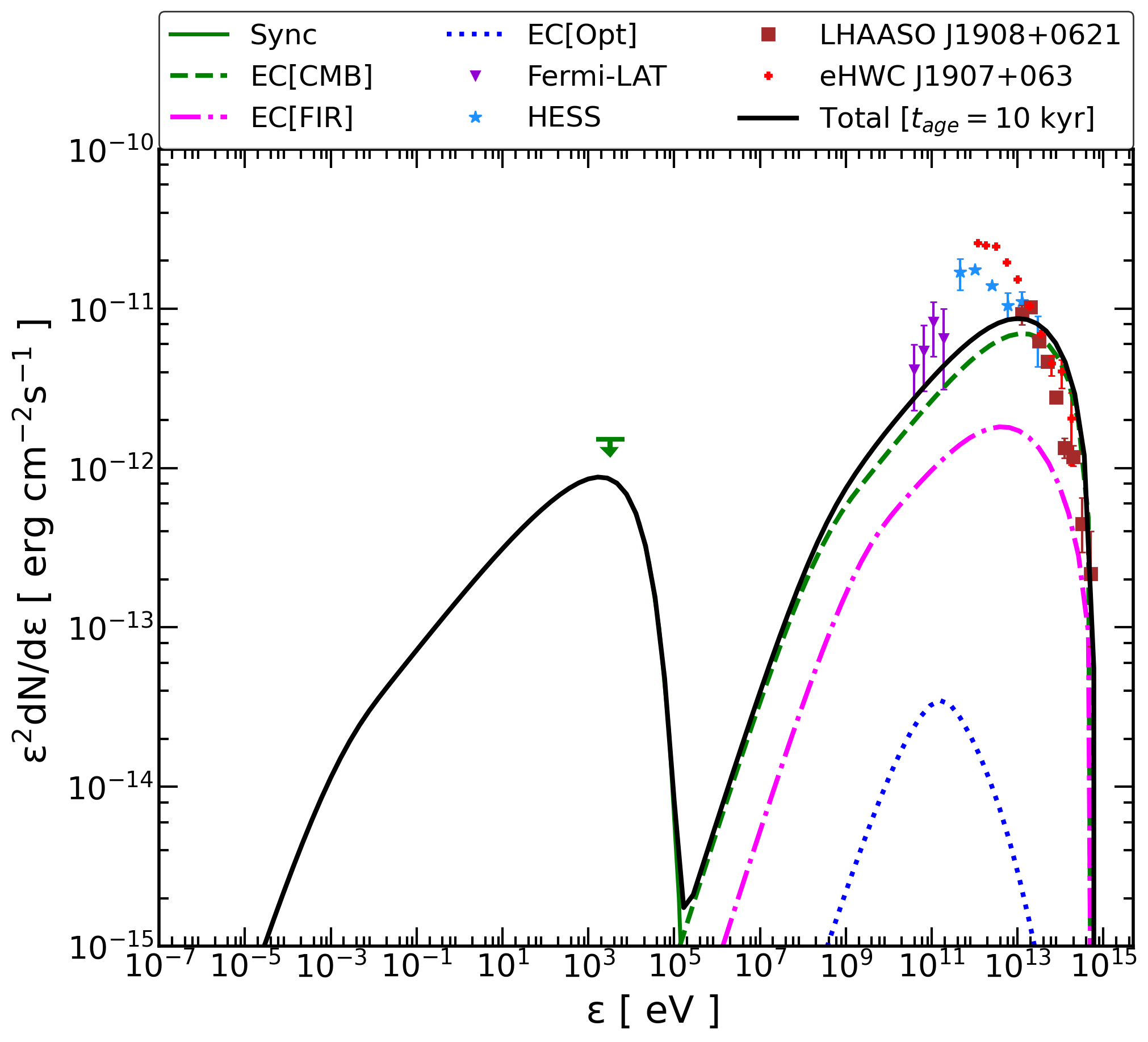}
\caption{{\em Top panel:} The temporal evolution of electron distribution in the PWN at different times for LHAASO J1908+0621. {\em Bottom panel:} The corresponding SED at $t_{\rm age} = 10$ kyr. The gamma-ray data points are taken from \protect\cite{cao2021Natur33C, bre5296B} and the X-ray upper limit is taken from \protect\cite{crestan2021MNRAS09C}.}
\label{fig:sed1908}
\end{figure}

The time-dependent non-thermal electron distributions and estimated SED at $t_{\rm age} = 10$ kyr are shown in the top and bottom panels of Figure \ref{fig:sed1908}. The model parameters are listed in Table \ref{parameters}. The electron spectrum inside the PWN is similar to a standard PWN and the maximum energy of the electrons is $\sim 734$ TeV, that is lower compared to the 1-zone model but higher compared to the 2-component model, as described in \cite{crestan2021MNRAS09C}. Our estimated value is different as this might be due to detailed modelling of the gamma-ray data. In their model, they found that a single accelerator is unable to interpret all the sets of observations. In that case, a 2-component model by \cite{crestan2021MNRAS09C} would be useful, however, our interpretation of the data is appropriate for our purpose of the estimation of the maximum energy of electrons inside the PWN.

\subsection{LHAASO J2226+6057}

LHAASO J2226+6057 is another Galactic UHE gamma-ray source reported by the LHAASO collaboration \citep{cao2021Natur33C}. It has a spatial association with SNR G106.3+2.7 and its PSR J2229+6114, which supports wind nebula (also called “Boomerang”) \citep{kothes236K_apj}. The MILAGRO collaboration also reported detections in the Boomerang PWN associated with PSR J2229+6114 \citep{abdo2009ApJ127A}. PSR J2229+6114 was a bright gamma-ray pulsar in the first FERMI-LAT catalog of gamma-ray pulsars \citep{abdo2010ApJS460A}. From the associated SNR, multi-TeV gamma-ray emissions were detected by the VERITAS detector named as VER J2227+608 \citep{acciar2009ApJ6A}. In the spatial region with VER J2227+608, Fermi-LAT detections observed GeV gamma-rays \citep{yuliangApJ162X}. The SNR G106.3+2.7 was detected in radio \citep{pine2000AJ8P} and also X-ray data is available for this source \citep{fujita2021ApJ1F}. 

We consider that PSR J2229+6114 with current spin-down luminosity $\dot{E} = 2.3 \times 10^{37}$ erg/s can power the PWN and UHE radiation from LHAASO J2226+6057. The distance to PSR J2229+6114 is $ \sim 3$ kpc based on X-ray absorptions \citep{halpern2001ApJ125H}. Note that the distance is uncertain and the atomic hydrogen and molecular velocity infers a distance $\sim 0.8$ kpc \citep{kothes236K_apj}. We have taken $d \sim 3$ kpc in our modelling of this source. The current period $P = 51.6$ ms and period derivative $\dot{P} = 78.3 \times 10^{-15}$ s/s implies the value of $\tau_c = 10.5 $ kyr \citep{abdo2010ApJS460A}. For PSR J2229+6114, we have assumed the age of PWN is $t_{\rm age} = 7$ kyr, similar to the value assumed in \cite{yu2022New9001669Y}. The source location is $(l, b) = [106.28^{\circ},2.83^{\circ}]$ or at $(R,z) = [9.7 , 0.15]$ kpc \citep{abdosecfc2013ApJ} in our Galaxy.

The origin of UHE gamma-rays has been tested based on the leptonic emission in PWN \citep{yu2022New9001669Y} and SNR G106.3+2.7 was investigated as a potential pevatron candidate by \cite{bao2021NatAs460T}. The GeV-TeV gamma-ray data of VER J2227+608 with a hard gamma-ray spectral index $ 1.90 \pm 0.04$ and a cutoff in the proton spectrum above 400 TeV indicate a plausible pevatron candidate \citep{yuliangApJ162X}.  In their PWN model \citep{yu2022New9001669Y}, they have used a power law type electron spectrum with a spectral index in the range of 2.3 to 2.5. In comparison, we have used a broken power law type electron distribution with spectral index 1.9 and 2.48, before and after the break energy.

\begin{figure}
\centering
%\subfloat[label 1]{{\includegraphics[width=7cm]{leptons.eps} }}%
\includegraphics[width=8.cm]{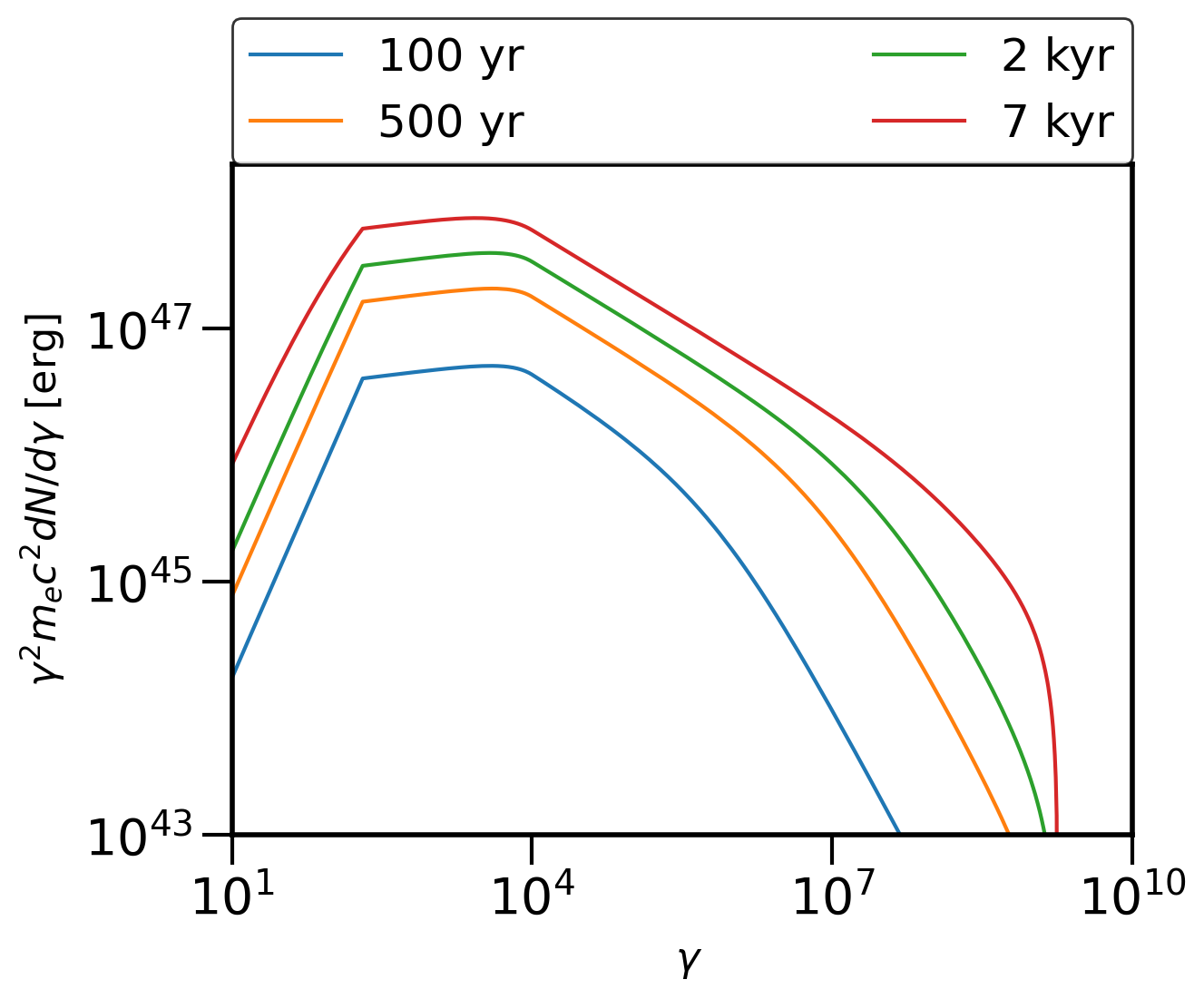}
\qquad
%\subfloat[label 2]{{\includegraphics[width=7cm]{lepton_pos.eps} }}%
\includegraphics[width=8 cm]{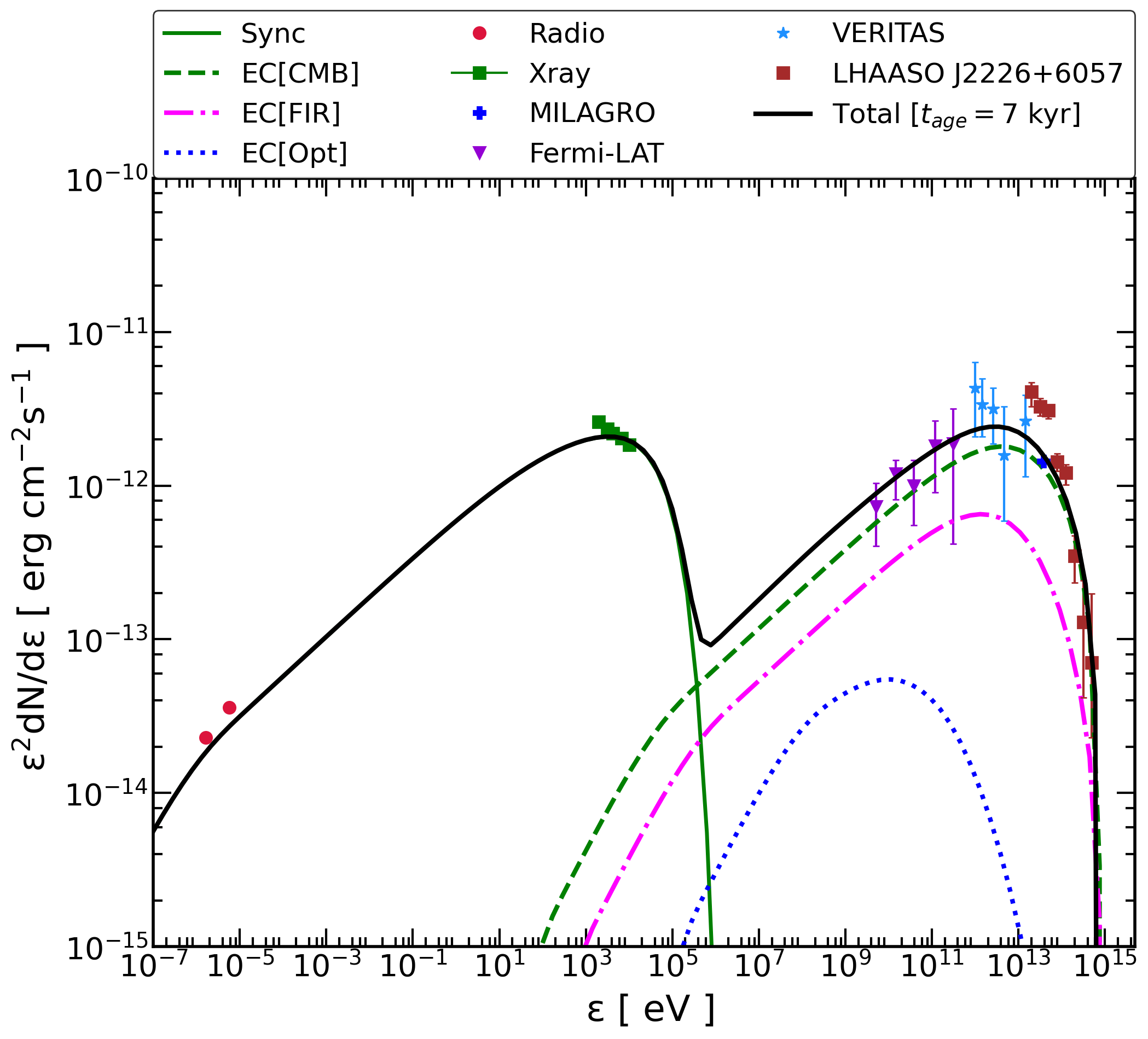}
\caption{{\em Top panel:} The temporal evolution of electron distribution in the PWN at different times for LHAASO J2226+6057. {\em Bottom panel:} The corresponding SED at $t_{\rm age} = 7$ kyr. The data points are taken from, Radio \citep{pine2000AJ8P}, X-ray \citep{fujita2021ApJ1F}, gamma-ray \citep{yuliangApJ162X,acciar2009ApJ6A,abdo2009ApJ127A, cao2021Natur33C,bre5296B}.}
\label{fig:sed2206}
\end{figure}

The time-dependent electron distribution at four epochs and SED at $t_{\rm age} = 7$ kyr are shown in the top and bottom panels of Figure \ref{fig:sed2206}, respectively. Our model constrains the average size of the PWN at the current age is approximately 11 pc, which is again in between the X-ray emitting and VHE/UHE gamma-ray emitting nebula. The emitting regions, constrained by the observations are $200^{''}$ in radio and X-rays \citep{halp2001ApJ23H, halpern2001ApJ125H}, the VHE size is $0.27^{\circ}$ \citep{acciar2009ApJ6A} and the UHE size of $0.49^{\circ}$ \citep{cao2021Natur33C}. The electron spectrum inside the PWN is similar to a standard PWN and the maximum energy of the electrons is $\sim 900$ TeV. For this source radio and X-ray observations are available and this provides better constraints on the PWN magnetic field. We find $B(t_{\rm age}) = 2.4 \mu$G for this source. In Figure \ref{fig:Pevatron}, we have fitted the SED with maximum electron energy $E_{\rm max} = 0.9, 1$ and 3 PeV respectively. Based on the UHE gamma-rays, we find that electrons of maximum energy 1 PeV are available inside this PWN. For 3 PeV, we find excess flux of UHE gamma-rays.

\subsection{HESS J1640-465}
This source was discovered by the HESS telescope during their survey of the inner Galaxy \citep{disc_2006ApJ777A}. \cite{slane2010ApJ66S,xin2018ApJ55X} found that the gamma-ray observations for this source can be explained by considering a PWN origin. Particle acceleration and radiation due to neutral pion decay in the SNR G338.3-0.0 can also account for the TeV emission in HESS J1640-465 \citep{abra2014MNRAS8A,tang2015Ap2T,2016AnAsupan, mares2021ApJ158M}. The TeV emission from HESS J1640-465 is spatially correlated with a SNR G338.3-0.0 and PSR J1640-4631 \citep{2016AnAsupan}. The {\it NuSTAR} X-ray observations of the pulsar enable the estimation of the braking index $n = 3.15 \pm 0.03$ for this source \citep{HBI2016ApJ16A}. However, for simplicity, we have used $n =3$, which is very close to the observed value. The characteristic age $\tau_c = 3113$ yr corresponding to the period $P = 206$ ms and $\dot{P} = 9.758 \times 10^{-13}$ s/s \citep{gott2014ApJ155G}, that infers a spin-down luminosity of $4.4 \times 10^{36}$ erg/s. In our calculations if we take the age of the pulsar or PWN the same as the SNR, which is found to be in between 5-8 kyr \citep{slane2010ApJ66S} then the spin-down timescale found to be negative based on equation \ref{eq_tage}, hence, we have chosen a lower value of $t_{\rm age} = 2.1$ kyr, that provides $\tau_0 \sim 1$ kyr. The distance to this source is in between 8.5-13 kpc based on the HI absorption \citep{Lemiere2009ApJ9L} and we have taken a reference value $d = 10$ kpc in our modelling.

Recently, \cite{mares2021ApJ158M} discussed the PWN origin of the multiwavelength observations of this source and they found a rapidly rotating pulsar with initial period $P_0 \sim 10$ ms is needed. They also required an extremely intense UV photon field, with energy density $1.3 ~{\rm keV/cm^3}$. HESS J1640-465 is a composite SNR source, and an young pulsar PSR J1640-4631 is also located in the shell-type SNR $\rm G338.3−0.0$ \citep{gott2014ApJ155G}.

\begin{figure}
\centering
%\subfloat[label 1]{a{\includegraphics[width=7cm]{leptons.eps} }}
\includegraphics[width=8.cm]{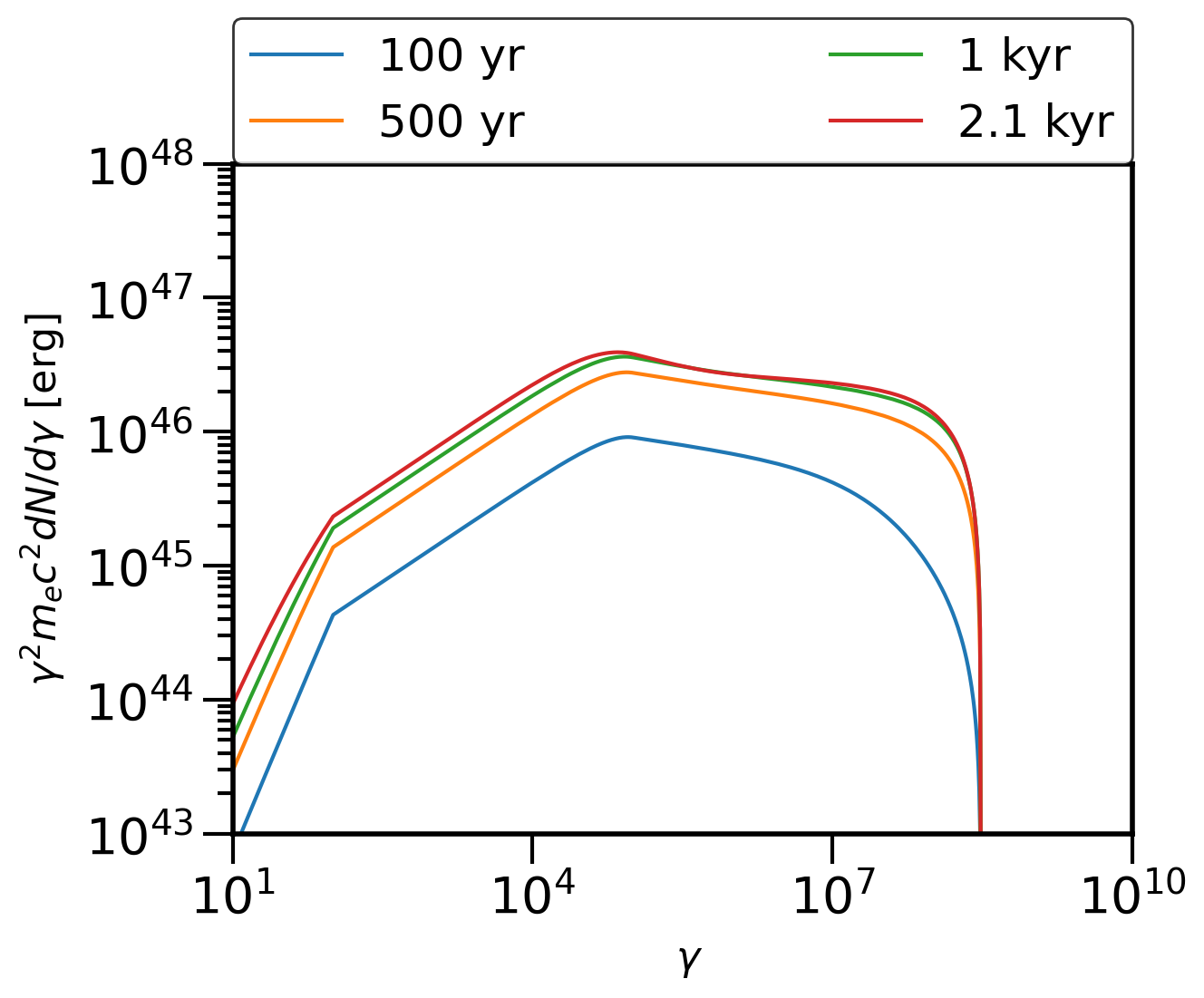}
\qquad
%\subfloat[label 2]{{\includegraphics[width=7cm]{lepton_pos.eps} }}%
\includegraphics[width=8 cm]{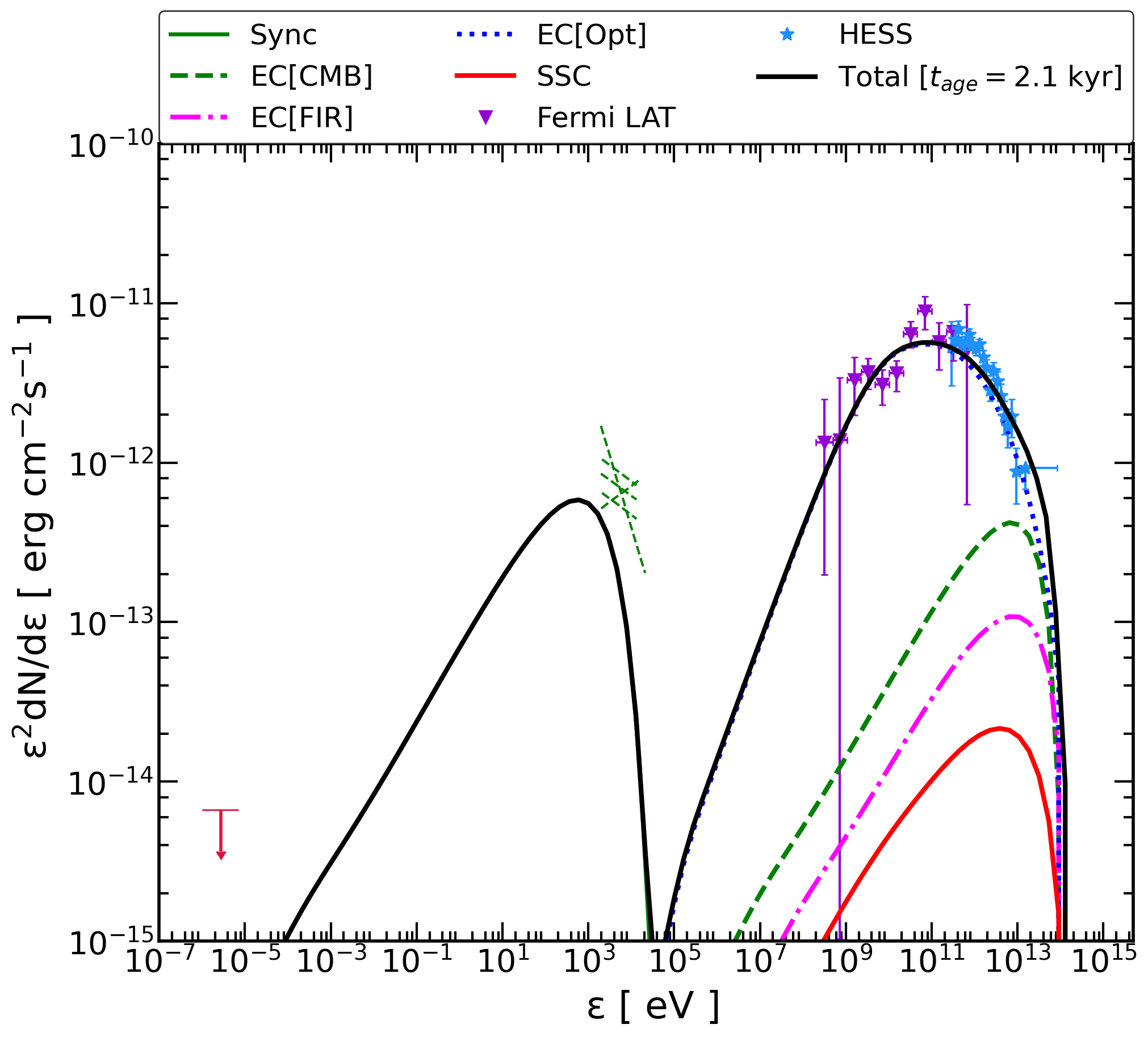}
\caption{{\em Top panel:} The temporal evolution of electron distribution in the PWN at different times for HESS J1640-465. The concave shape of the electron distribution at late times is due to the dominant IC cooling in the dense stellar photon environment. {\em Bottom panel:} The corresponding SED at $t_{\rm age} = 2.1$ kyr. The radio flux upper limits and the remaining broadband data are extracted from \protect\cite{abra2014MNRAS8A, mares2021ApJ158M}.}
\label{fig:1640} 
\end{figure}

As also mentioned by \cite{mares2021ApJ158M} about its closeness with a nearby young massive stellar cluster, which is $8^{'}$ away from HESS J1640-465 \citep{davies2012MNRAS}. This cluster can provide a higher stellar photon density at the pulsar location $(l, b) = [338.28^{\circ}, -0.04^{\circ}]$ or at $(R,z) = [3.78, -0.005]$ kpc \citep{abdalla2018AnA1H}. We also found that for this source stellar photon density 1.3 ${\rm keV/cm^3}$ is required. Further, \cite{mares2021ApJ158M} has taken the age of the pulsar equal to 3 kyr, which is a slightly larger value compared to our modelled value $t_{\rm age} = 2.1$ kyr. This creates a difference in the spin-down timescale. The magnetization parameter proposed by \cite{mares2021ApJ158M}, $\eta_B = 10^{-1}$  is very large compared to our modeled value  $\eta_B = 10^{-4}$ and this is also contrary to the convention that PWNe are particle dominated. Our used value of gas density $0.1 ~\rm cm^{-3}$ is 10 times higher comparatively and the spin-down timescale is larger by a factor of 250. The diversity in the model parameters for the interpretation of the PWN emission is evident between these models and infers the degeneracy in input model parameters. The electron distribution at four epochs and the SED at $t_{\rm age} = 2.1$ kyr are shown in the top and bottom panels of Figure \ref{fig:1640}, respectively. At this age, the modelled value of the PWN radius is approximately 2.1 pc which is a factor of 2 lower compared to the observed Xray size of the nebula $3.5$ \citep{gott2014ApJ155G} and approximately 10 times lower compared to HESS measured size \citep{abram2014ApJss}. The maximum energy of electrons based on the spectral fit is $\sim$ 137 TeV.

\subsection{HESS J1813-178}
HESS J1813-178 was discovered during the HESS survey of Galactic sources \citep{ahar2005Sci1938A,disc_2006ApJ777A}. Its multiwavelength observations are available in radio, X-ray \citep{brogan2005ApJ5B}, and gamma-rays \citep{araya2018ApJ9A}. This source has spatial overlap with the SNR G12.82-0.02 and at the centre of the SNR, PSR J1813-1749 is located \citep{gott2014ApJ155G}. The current spin-down luminosity of the pulsar is $\dot{E} = (6.8 \pm 2.7) \times 10^{37}$ erg/s \citep{gott2009ApJG,camilo2021ApJ67C}. Using the pulsar period $P = 44.7$ ms and its derivative $\dot{P} = 1.265 \times 10^{-13}$ s/s, we estimated $\tau_c = 5.6$ kyr \citep{halpern2012ApJ14H}. 

\begin{figure}
\centering
%\subfloat[label 1]{a{\includegraphics[width=7cm]{leptons.eps} }}
\includegraphics[width=8.cm]{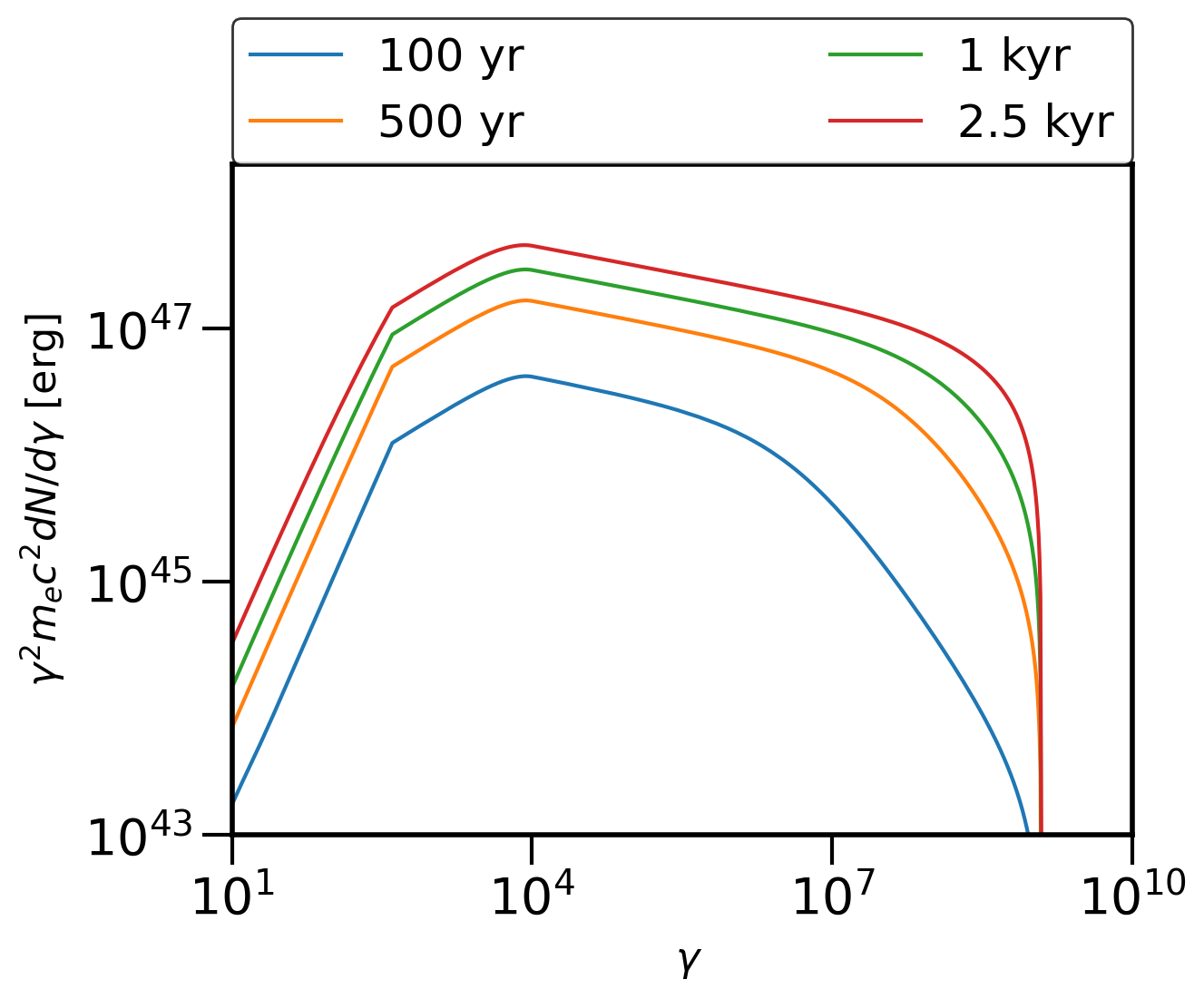}
\qquad
%\subfloat[label 2]{{\includegraphics[width=7cm]{lepton_pos.eps} }}%
\includegraphics[width=8 cm]{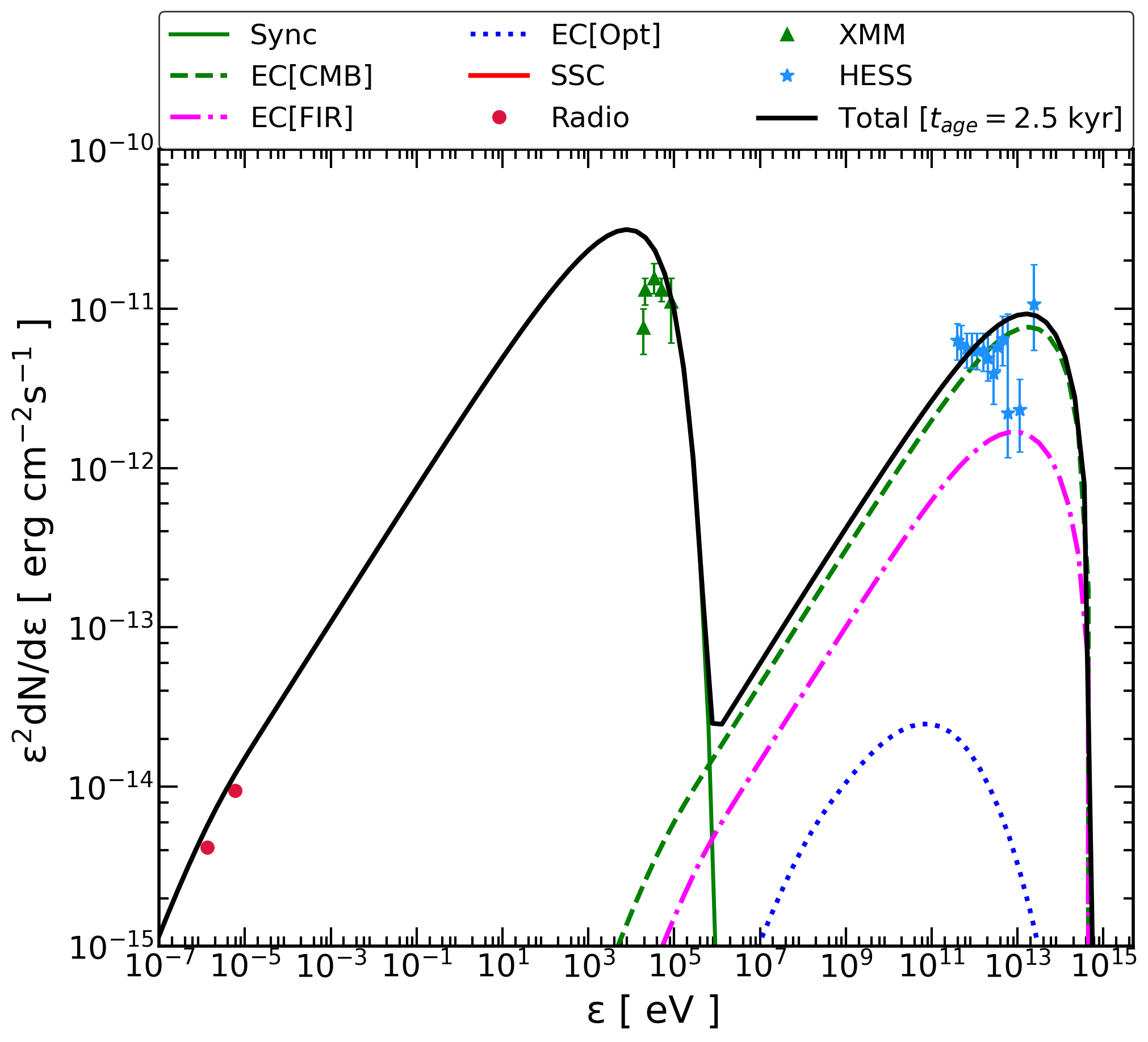}
\caption{{\em Top panel:} The temporal evolution of electron distribution in the PWN at different times for HESS J1813-178. {\em Bottom panel:}  The corresponding SED at $t_{\rm age} = 2.5$ kyr. The MW data for HESS J1813-178 is taken from \citep{fang2010ApJ7F}.}
\label{fig:4} 
\end{figure}

The radio observations of $\rm G12.82−0.02$ infers a compact size $3^{'}$ and also, independent of distance, the age of the SNR is between 285-2,500 yrs \citep{brogan2005ApJ5B}. The distance of the source is quite uncertain and a value of 4.7 kpc was used to study the non-thermal radiation \citep{fang2010ApJ7F}. In recent studies, a large dispersion measure was found for the pulsar and, based on the electron distribution model of \cite{cordes2002astrC}, the source distance was found to be $12 \pm 2$ kpc. We caution that the dispersion in the distance will affect the model parameters. We have taken its updated distance of 6.2 kpc based on the electron distribution model of \cite{yao2017ApJ29Y}. The source location is $(l, b) = [12.81^{\circ}, -0.03^{\circ}]$ or at $(R,z) = [2.81 , -0.0032$] kpc in our Galaxy \citep{disc_2006ApJ777A}. The electron distribution at four epochs and the SED at $t_{\rm age} = 2.5$ kyr are shown in the top and bottom panels of Figure \ref{fig:4}, respectively. The emitting regions for this source in the X-ray band infer a size of value 2.4 pc \citep{uber2005ApJ109U,funk2007AnAF,helfand1297H}, while the radio region is of size 5.4 pc \citep{brogan2005ApJ5B}. Also, in the case of this source, the HESS observations infer a very compact size (4 pc) of the emission region \citep{disc_2006ApJ777A}. Our modelled value of the PWN radius is approximately 4 pc. Further, for this source radio and X-ray observations are available and this provides better constraints on the PWN magnetic field. We find $B(t_{\rm age}) = 4 \mu$G for this source. The maximum energy of the electrons is $\sim 628$ TeV.

\subsection{HESS J1303-631}

\begin{figure}
\centering
%\subfloat[label 1]{a{\includegraphics[width=7cm]{leptons.eps} }}
\includegraphics[width=8.cm]{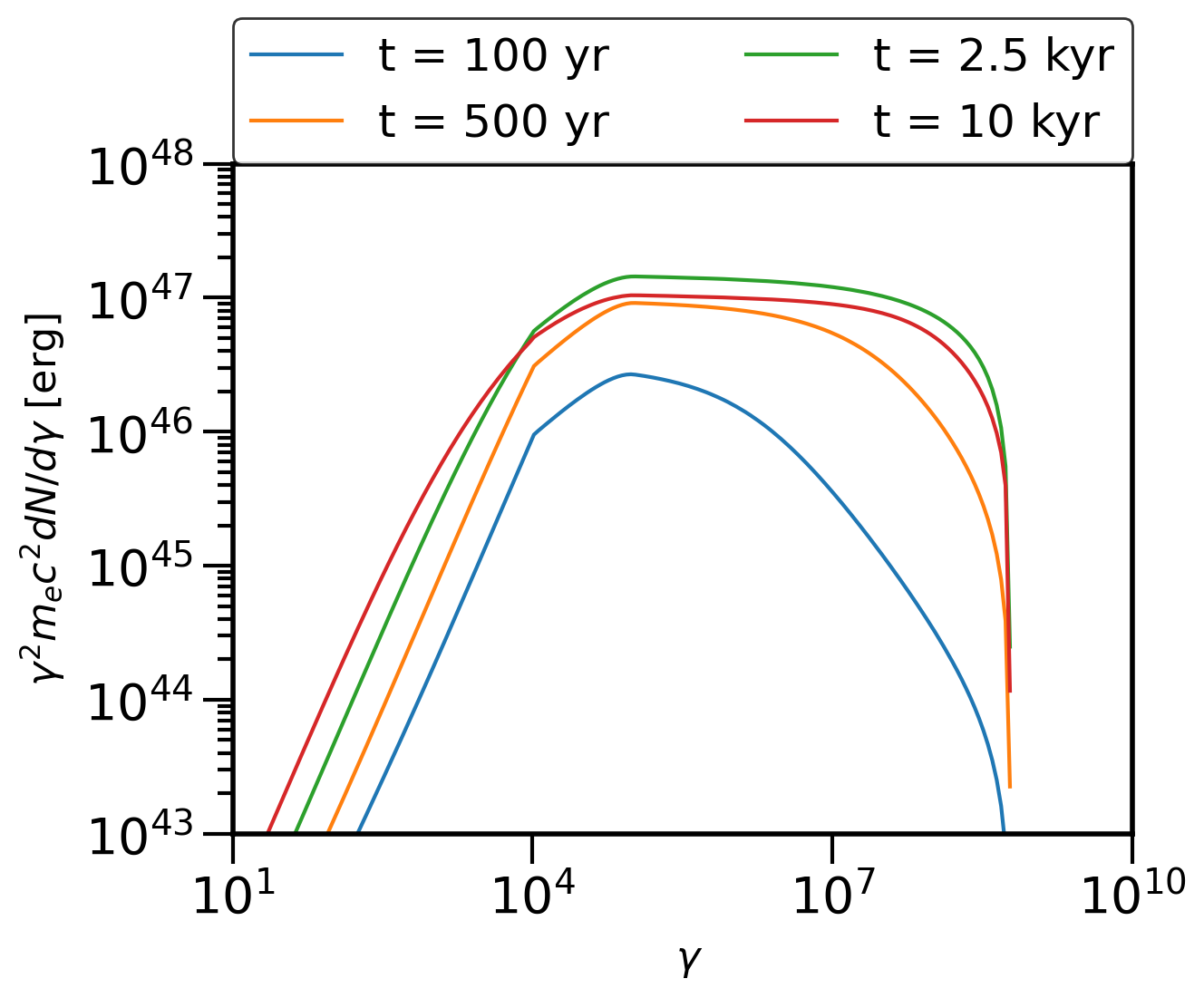}
\qquad
%\subfloat[label 2]{{\includegraphics[width=7cm]{lepton_pos.eps} }}%
\includegraphics[width=8 cm]{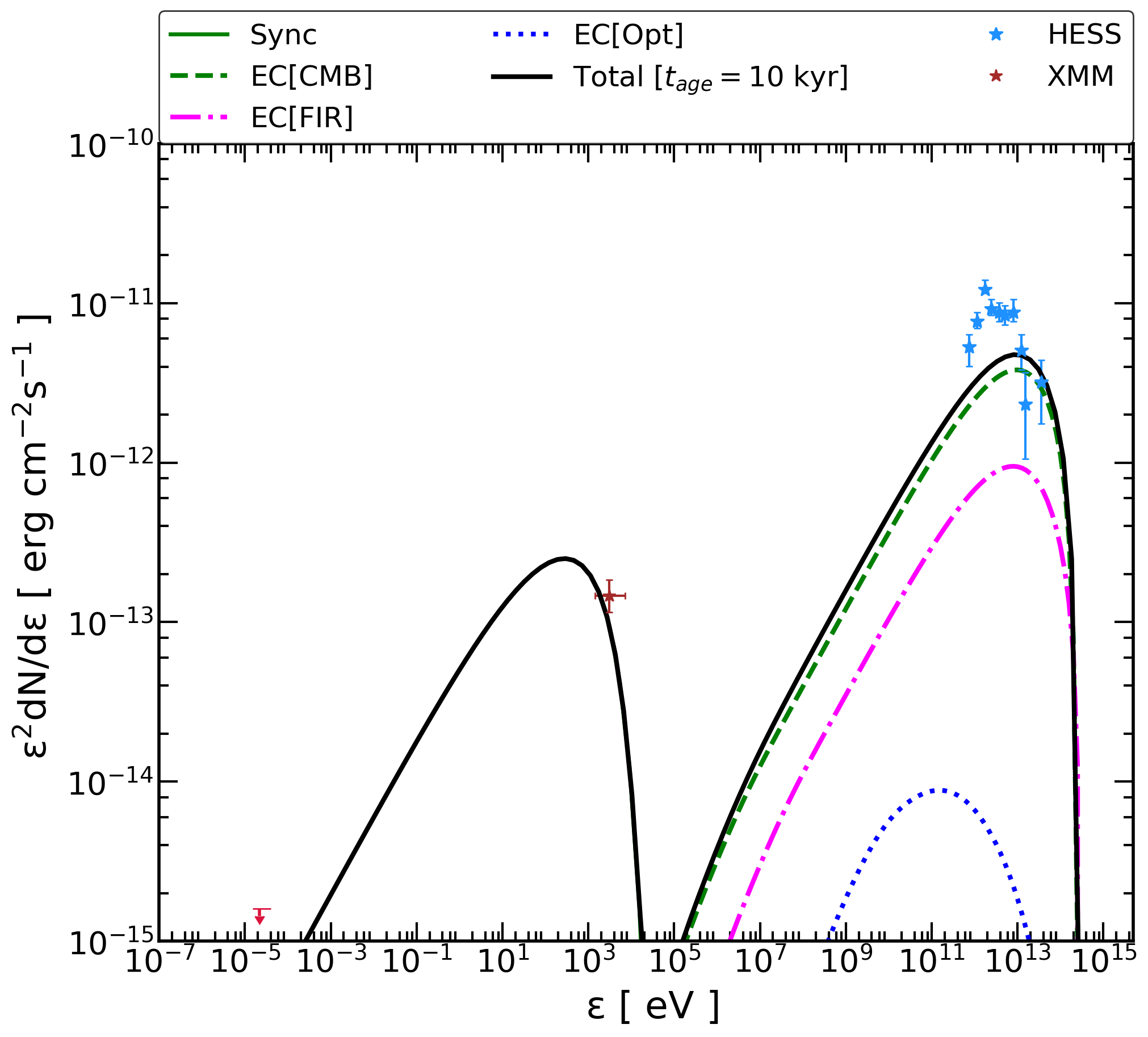}
\caption{{\em Top panel:} The temporal evolution of electron distribution in the PWN at different times for HESS J1303-631. {\em Bottom panel:} The corresponding SED at $t_{\rm age} = 10$ kyr. The upper limits for the radio, X-ray data, and GeV-TeV data of HESS J1303-631 are taken from \citep{hess2012AnA46H}.}
\label{fig:sed1303} 
\end{figure}

The multiwavelength (radio, X-ray, and gamma-ray) observations of HESS J1303-631 can be explained through the PWN emission \citep{hess2012AnA46H}. They have used a stationary leptonic model with magnetic field value 1.4 $\mu G$. In our revised version, the magnetic field is lower and in our modelling we evolve the electron population under the adiabatic and cooling losses plus reverberation and interpret the data. The distance to this source is 6.6 kpc based on the Galactic electron distribution model by \cite{cordes2002astrC} and the IR energy density is $1.3~{\rm eV/cm^3}$ at the location $(l,b) = [304.21^{\circ}, -0.33^{\circ}]$ or at $(R,z) = [7.25, -0.04]$ kpc \citep{disc_2006ApJ777A}. We revisit the source modelling using time-dependent PWN emission. The source is associated with PSR J1301-6305 having period $P= 184$ ms, period derivative $\dot{P}  = 2.65 \times 10^{-13}$ s/s and spin-down luminosity $\dot{E} = 1.7 \times 10^{36}$ erg/s \citep{manches2005AJ129M,hess2012AnA46H}. This provides the characteristic age of this source to be approximately 11 kyr. The electron distribution at four epochs and the SED are shown in the top and bottom panel of Figure \ref{fig:sed1303} for $t_{\rm age} = 10$ kyr. There is no radio counterpart for this source \citep{sushch2017AnAS} but the X-ray size is of value $\sim 4$ pc \citep{hess2012AnA46H} and a larger emitting region in the VHE gamma-rays of value 18.4 pc \citep{ahar2005AnA3A}. Our model value for the PWN radius is 13.8 pc and it's closer to the VHE gamma-ray emitting nebula. For our selected input parameters the full part of the VHE gamma-ray spectrum is not explained however, it was explained by \citet{hess2012AnA46H}. This difference might be due to the contrast difference in our modelling and IR field $> 0.2 ~\rm eV/cm^3$ can be useful for reproducing the full VHE spectrum. However, we focus on the estimation of the maximum energy of the electrons based on the maximum energy of the VHE photons. The maximum energy of the electrons from the spectral fit is $\sim 307$ TeV.

\section{Discussions and Conclusions}

In recent years, the VHE gamma-ray spectrum of Galactic sources has been extended to the UHE range by the HAWC and LHAASO detectors. The discovery of 1.1 PeV photon from the Crab PWN makes it one of the first Galactic sources of PeV photons and also a Galactic leptonic  pevatron source \citep{crpevlhaso}. \cite{2021arXiv210914148L} have shown that the end part of the UHE gamma-ray spectrum is harder for the CRAB Nebula and infers the acceleration of cosmic ray (CR) protons up to PeV energies. Motivated by these results we have explored the origin of the UHE gamma-ray spectrum in two of the LHAASO detected sources. We have taken the cooling of the electrons in the KN regime \citep{mode2005MNRAS.363..954M} while calculating the electron distribution and in the SED calculation these modifications are already present in the NAIMA public code based on the formalism by \cite{khangul2014ApJK}. Further, we compare these model parameters of two LHAASO detected PWN with a few other H.E.S.S. detected objects studied by us. We find the UHE detected objects can be interpreted using the spin-down luminosity of pulsars and their model parameters are similar to modelled VHE detected sources. Hence, most of the VHE gamma-ray sources in our Galaxy are powered by the pulsar spin-down luminosity then they should also be detected by UHE gamma-ray detectors. Recently, this is also shown by \cite{albert2021ApJ27A}, that the powerful pulsars with $\dot{E} \ge 10^{36}$ erg/s, would have a UHE gamma-ray spectrum.

\begin{figure}
\centering
%\subfloat[label 1]{{\includegraphics[width=7cm]{leptons.eps} }}

%\includegraphics[width=7.cm]{elpop_1908.png}
\qquad
%\subfloat[label 2]{{\includegraphics[width=7cm]{lepton_pos.eps} }}%
\includegraphics[width=9 cm]{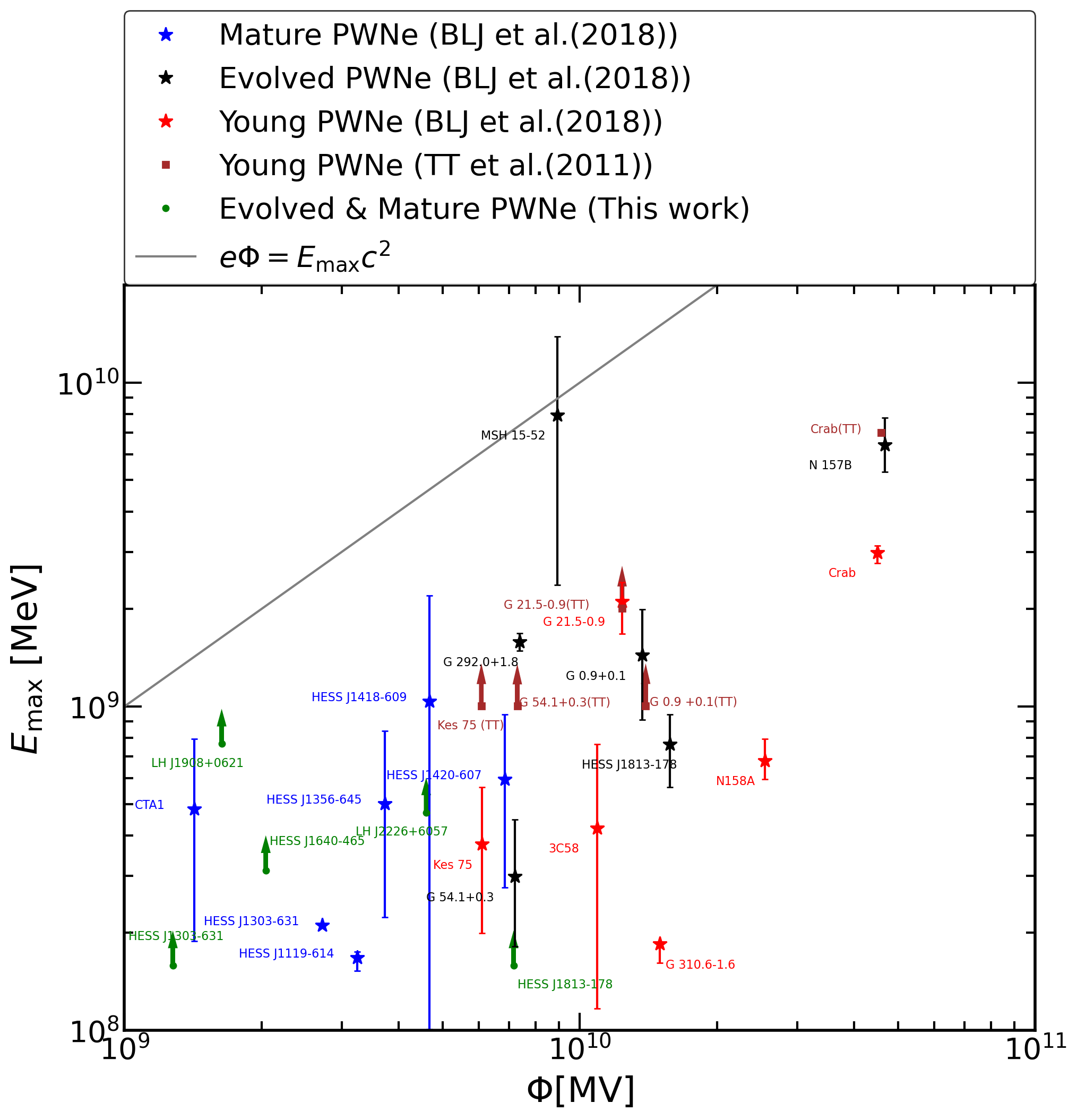}
\caption{The multiwavelength fit based maximum energy $E_{\rm max}$ vs polar cap potential $\unslant\phi$. We have also included these scaling values from \citep{2018zhu609A110Z} (BLJ et al.(2018)) and \citep{tanaka2011ApJT} (TT et al.(2011)) for the comparison.The young PWNe have age less than 2.5 kyr,  and for evolved PWNe the age is above 2.5 kyr but less than 5 kyr and for mature PWNe it is above 5 kyr respectively as defined in \citep{2018zhu609A110Z}.}
\label{fig:gmaxval}
\end{figure}

In Table \ref{parameters}, we have listed parameters: (i) From Past Observations (ii) Assumed parameters (iii) Parameters based on spectral fit (iv) Derived from Parameters in (i) and (ii) and (iii). Our current estimation of the PWN radius can't interpret the wavelength dependent size of the emitting nebula. The variation of size in radio to gamma-ray wavelengths is very explicit from observations. We think that age is also an important factor in deciding the radius of the PWN. Wavelength dependent effects on PWN size requires the inclusion of the particle particle transport scenarios \citep{2012ApJ75283T}.  The evolution of the PWN radius during the ejecta and ST phase is shown in Figure \ref{fig:radevl}. The injected electron spectrum used in our study follows a broken power law. The spectral index $p_1$ and $p_2$ are consistent with the standard PWN interpretation. Due to the very weak magnetic field at the current age of the PWN, the cooling Lorentz factor is lower than the minimum Lorentz factor for all the sources. For the breaks in the injected electron spectrum, we can find their signatures in their SEDs. The magnetic field inside LHAASO J1908+0621 is 0.55 $\mu$G and for HESS J1303-631 is 0.5 $\mu$ G and the origin of these low magnetic field values inside PWN is not very well known. In our case, these two objects are older, compared to others and that leads to these small values in the expansion phase. However, similar low values have been reported for the PWN modelling, for example, in the modelling of UHE gamma-ray source, HAWC J1826−128 by \cite{2022ApJ...930..148B}.

\begin{figure}
\centering
%\subfloat[label 1]{{\includegraphics[width=7cm]{leptons.eps} }}

%\includegraphics[width=7.cm]{elpop_1908.png}
\qquad
%\subfloat[label 2]{{\includegraphics[width=7cm]{lepton_pos.eps} }}%
\includegraphics[width=9 cm]{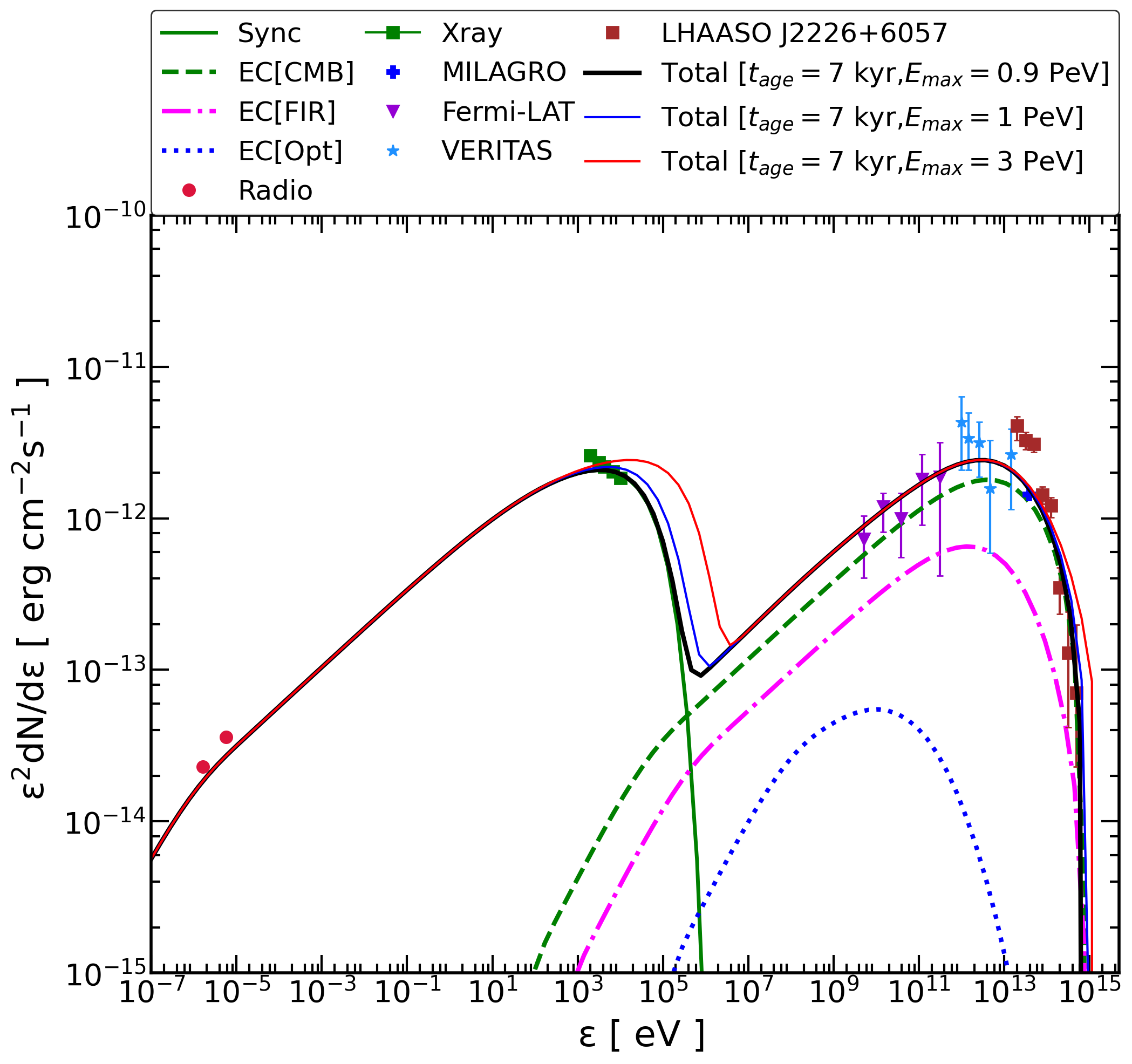}
\caption{The SED of LHAASO J2226+6057 is shown for maximum electron energy $E_{\rm max} = 0.9, 1$ and 3 PeV, respectively. The other parameters are the same as listed in Table \ref{parameters} for this source. The UHE gamma-rays infer that electrons of PeV energy can be inside the PWN.}
\label{fig:Pevatron}
\end{figure}
The value of $\eta_e$ is dominant compared to $\eta_B$ for all sources and implies that the PWN plasma is dominated by the $e^{\pm}$ pair-plasma. Similar conclusions about the PWN composition were reported in earlier studies \citep{2013ApJ763L4T}. Further, using fitted parameters, we have estimated the average energy per particle in the pulsar wind $\Gamma_w$ and the number of pairs produced per photon, i.e., pair multiplicity $\kappa$. The values of $\kappa$ are approximately in between $\sim 10^4-10^7$.

Our assumed values for external radiation field energy density for the IR and stellar photons are within the standard values as known from the Galactic radiation field models \citep{pop2017MNRAS,bre5296B,2018zhu609A110Z}, however, these values can be location dependent. Surprisingly, For the HESS J1640-465 source, a very large photon density is supported by a nearby source as discussed in Section 3.  Further, the temperature of the IR radiation field can affect the IC radiation and lower values are useful to produce UHE radiation \citep{2021ApJ908L49B}, we have taken $T_{\rm IR} = 20$ K in this work and this value is consistent with the dust temperature \citep{bernard2010AnA8B, zhu2014AnA111Z}.  With more UHE source detection in the future by LHAASO and CTA, the IR radiation environments can be tested.

The $\gamma-\gamma$ absorption in the ISM can affect the VHE to UHE part of the gamma-ray spectrum \citep{2006moska_att}. However, for the two LHAASO objects these effects on the TeV-PeV gamma-ray spectrum are negligible \citep{cao2021Natur33C}. For HESS J1813-178 and HESS J1303-631, we found that attenuation is not important. In the case of HESS J1640-465, these effects are dominant due to the high density of the target photons. To minimize it, we have reduced the target photon temperature to 5000 K and an exponential cut-off in the gamma-ray spectrum is used above $\sim$ 50 TeV based on the pair production condition $\epsilon_o \epsilon_{\gamma} \ge (2m_e c^2)^2$, where $\epsilon_o$ and $\epsilon_{\gamma}$ are the energy of optical and gamma-ray photons, respectively.

\begin{figure}
\centering
%\subfloat[label 1]{a{\includegraphics[width=7cm]{leptons.eps} }}
\includegraphics[width=7.cm]{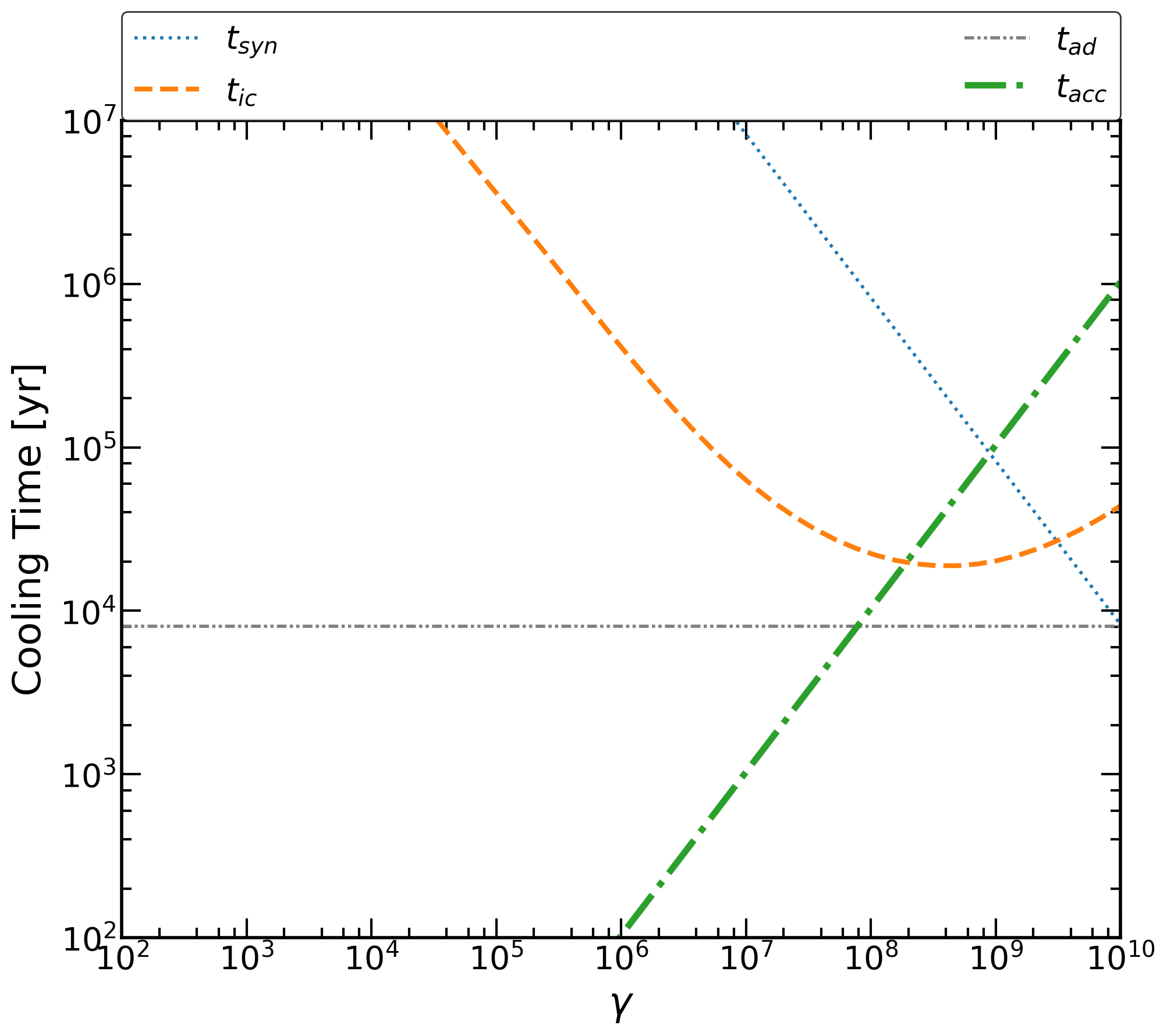}
\qquad
%\subfloat[label 2]{{\includegraphics[width=7cm]{lepton_pos.eps} }}%
\includegraphics[width=7 cm]{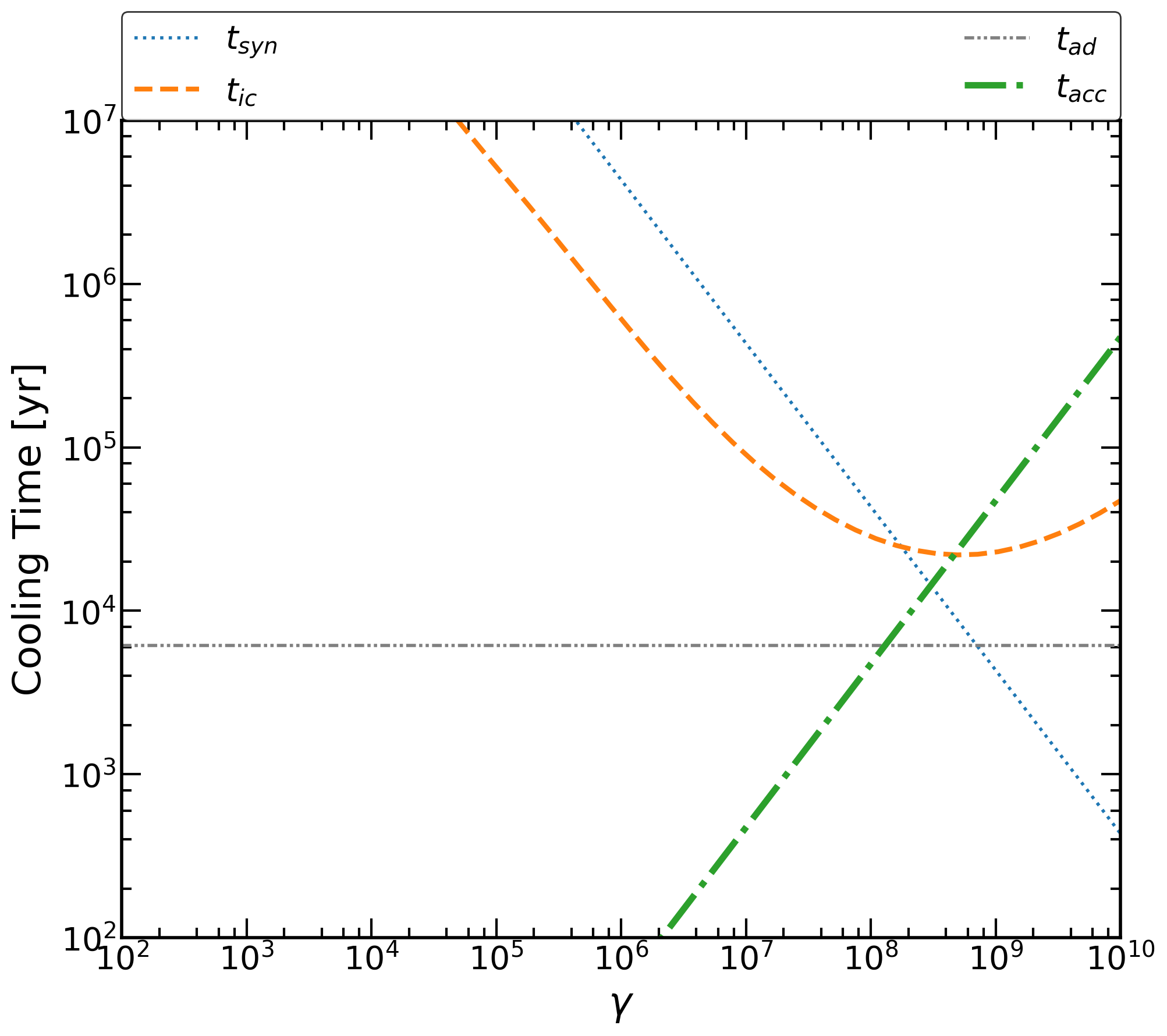}
\caption{LHAASO J1908+0621 and LHAASO J2226+6057 cooling timescales for the model parameters listed in Table \ref{parameters}.}
\label{tA} 
\end{figure}

\begin{figure}
\centering
%\subfloat[label 1]{a{\includegraphics[width=7cm]{leptons.eps} }}
\includegraphics[width=7.cm]{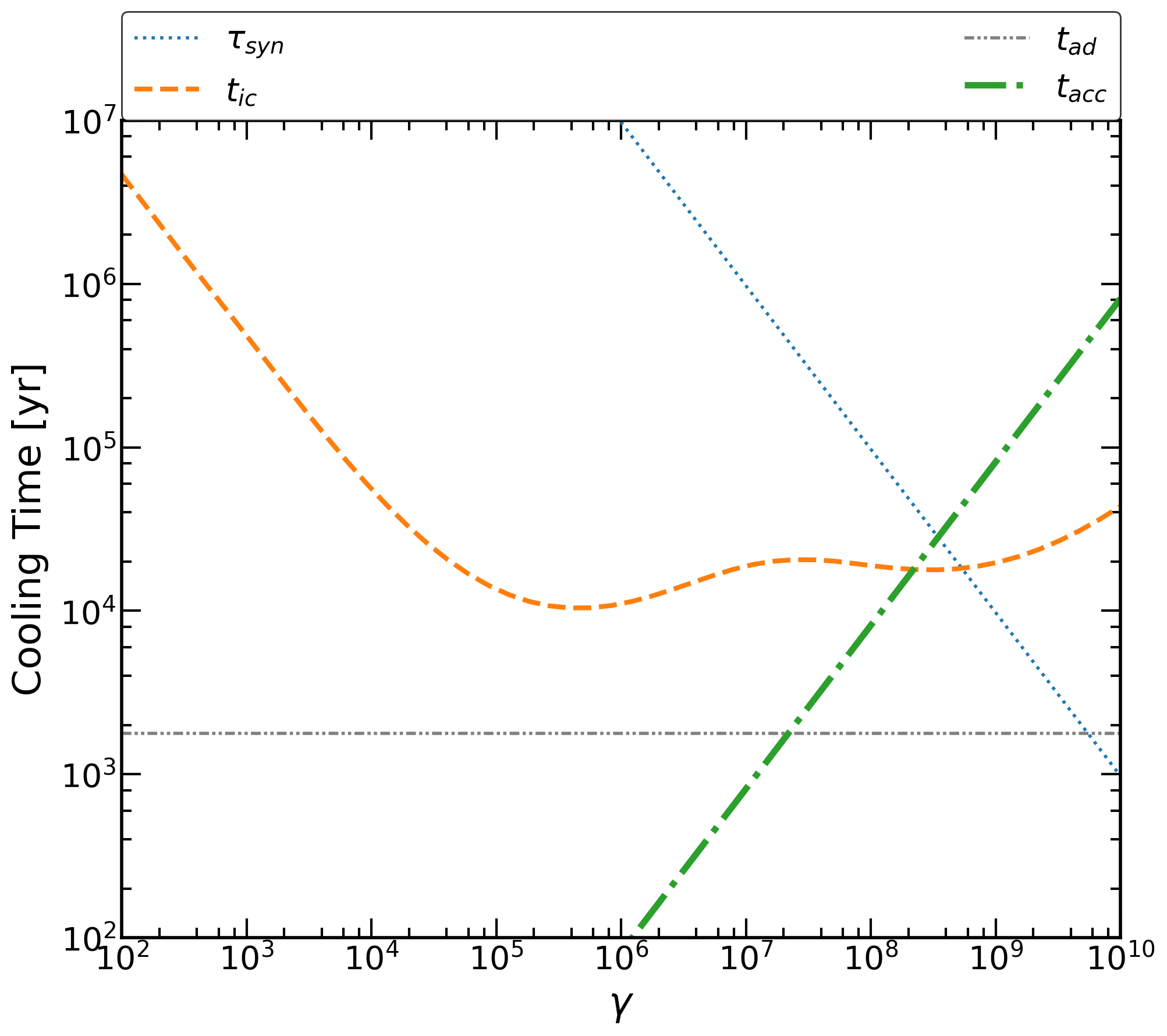}
\qquad
%\subfloat[label 2]{{\includegraphics[width=7cm]{lepton_pos.eps} }}%
\includegraphics[width=7 cm]{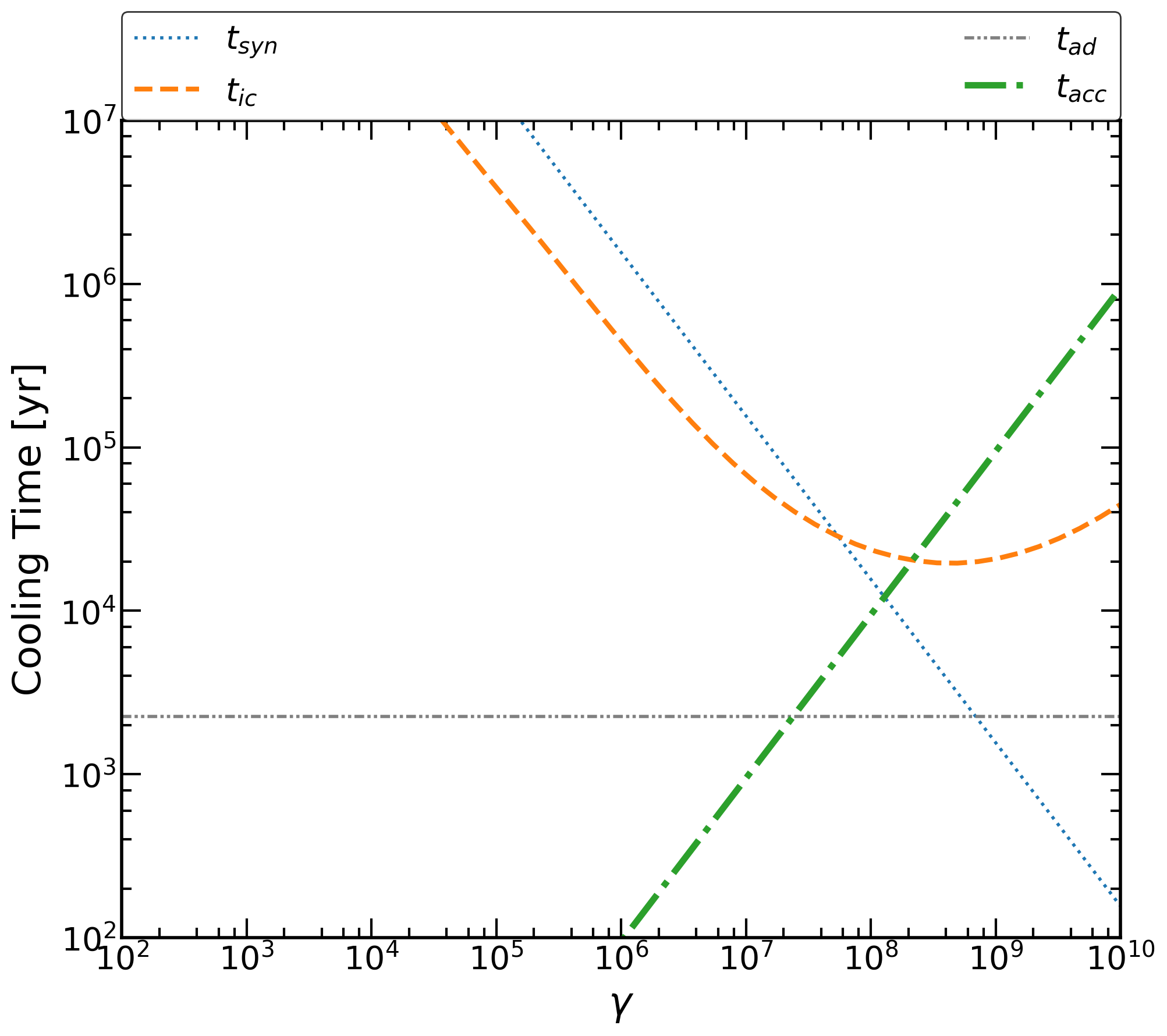}
\caption{HESS J1640-465 and HESS J1813-178 cooling timescales for the model parameters listed in Table \ref{parameters}.}
\label{tB} 
\end{figure}

\begin{figure}
\centering
%\subfloat[label 1]{{\includegraphics[width=7cm]{leptons.eps} }}

%\includegraphics[width=7.cm]{elpop_1908.png}
\qquad
%\subfloat[label 2]{{\includegraphics[width=7cm]{lepton_pos.eps} }}%
\includegraphics[width=7 cm]{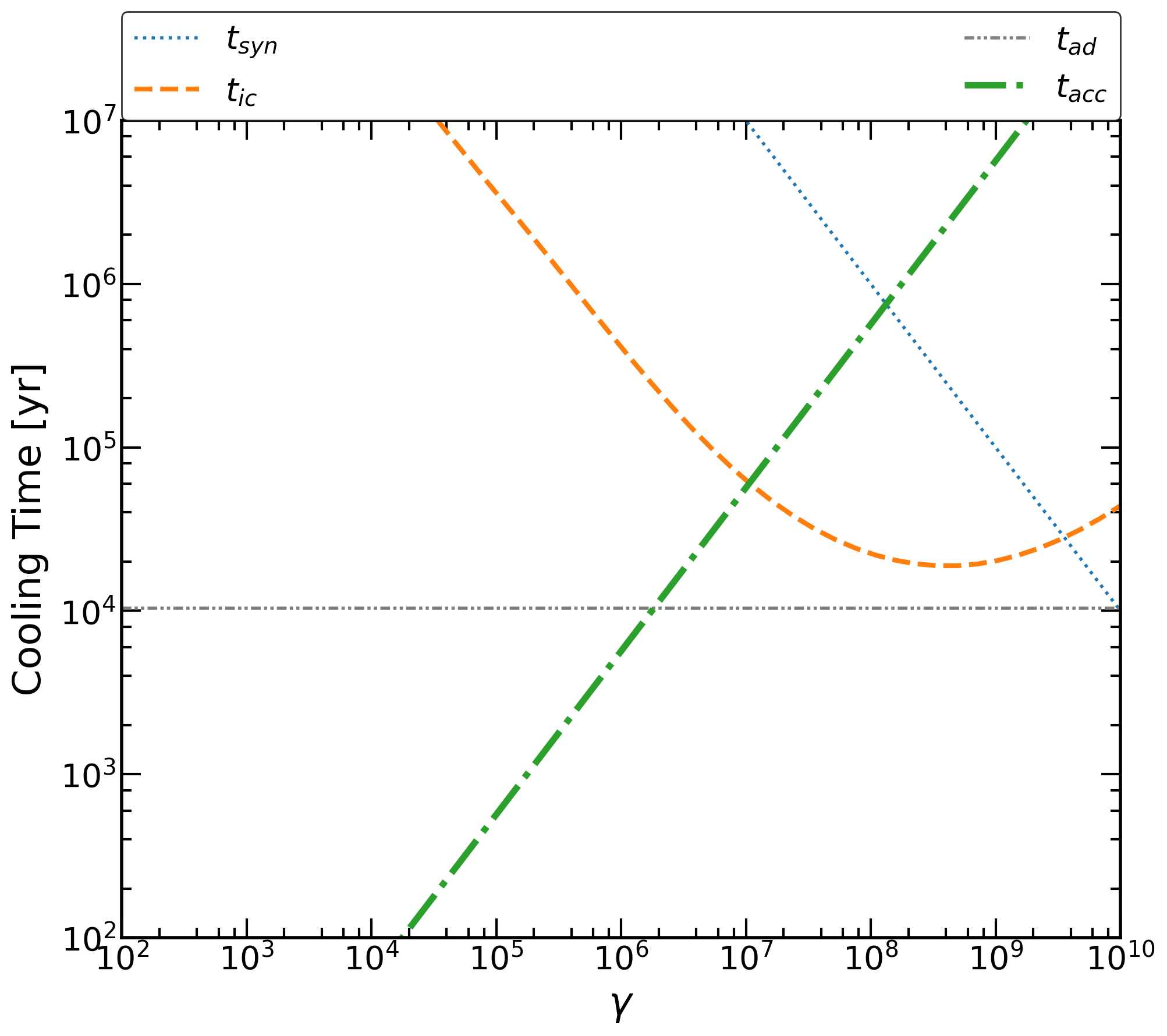}
\caption{HESS J1303-631 cooling timescales for the model parameters listed in Table \ref{parameters}.}
\label{tC}
\end{figure}

Using the maximum synchrotron photon energy we calculate acceleration efficiency $\eta_{\rm acc}$ of the electrons and we found its values in between $10^{-4}-10^{-5}$, also listed in Table \ref{parameters}.  Using, equation 11, we obtained values of $\gamma_{\rm max, cool}$. These values are approximately 10 times lower than $\gamma_{\rm max, PC}$. The required values of $\gamma_{\rm max}$ from the spectral fit are in between $\gamma_{\rm max, cool}$ and $\gamma_{\rm max, PC}$. Hence, the maximum energy of the electrons is supported by the polar cap potential \citep{deo_wilhelm_2022}. In Figure \ref{fig:gmaxval}, we have shown the scaling of the maximum energy of electrons based on the multiwavelength fit, vs the polar cap potential $\unslant\phi = E_{\rm max, PC}/e$.  The cooling timescales for particles inside PWN are shown in Figure \ref{tA}-\ref{tC}, for the current age of the pulsar and input model parameters. It is clear that within the acceleration timescale, adiabatic and synchrotron cooling mechanisms dominate and limit the maximum energy of electrons at the termination shock. To interpret the multiwavelength radiation we need another source of electrons above these energies and hence, the injection of maximum energy particles inside PWN  must be due to the polar cap potential regions. Our estimated values of the maximum electron energy from the spectral fit is in the range 0.1-1 PeV. In Figure \ref{fig:Pevatron}, we have shown the SED modelling of LHAASO J2226+6057 at 1 and 3 PeV. We have selected it as for this object radio, X-ray, and UHE observations are available). This shows that UHE gamma-ray spectrum is a key to probing the PeV electron presence inside the PWN and makes LHAASO J2226+6057 one of the potential leptonic pevatron candidates compared to other sources investigated in this work.

\section{Acknowledgements}

JCJ is thankful to J. Carlos, N. Jingade, P. Kushwaha, R. Liu, V. Chand, E. Amato, N. Gupta, and X.-Y. Wang for helpful discussions and S. Crestan for providing the multiwavelength data files for the source MGRO J1908+06. SJT would like special thanks to T. Tezuka for providing the base of the numerical code used in this study. SJT is supported by Aoyama Gakuin University-Supported Program “Early Eagle Program”. SR was supported by a grant from the University of Johannesburg Research Council. Further, we are very thankful to A. Joshi, N. Fraija and M. Cardillo for reading our work and for insightful comments. 

\section{DATA AVAILABILITY}

The data used in this work are available in the article and the code developed for this work
can be shared on reasonable request to the corresponding author.

%\onecolumn

\footnotesize{
\bibliography{ref_prop}
}

%\onecolumn

%\section*{Appendix A}
%\subsection*{PWN radius evolution inside SNR}

%Based on \citep{gelfand2009ApJ2051G}, we have calculated the dynamical evolution of the PWN radius.

\end{document}